\documentclass[twocolumn,prd,nofootinbib,superscriptaddress,eqsecnum,tightenlines,8pt]{revtex4}

\usepackage{fancyhdr}
\usepackage{amsmath,amsfonts,amssymb}
\usepackage[colorlinks,citecolor=blue,linkcolor=blue,urlcolor=blue]{hyperref}
\usepackage[english]{babel}

\usepackage{color}
\definecolor{red}{rgb}{1,0,0}
\definecolor{gre}{rgb}{0,0.6,0}
\definecolor{blu}{rgb}{0,0,1}

\usepackage{mathenv}
\usepackage{ulem}

\usepackage[pdftex]{graphicx}
\usepackage{mathrsfs} 
\DeclareGraphicsExtensions{.jpg, .JPG , .png , .pdf, .gif}
\def\be{\begin{equation}}
\def\ee{\end{equation}}

\begin{document}

\title{Scalar spectra of primordial perturbations in loop quantum cosmology}


\author{Aur\'elien Barrau}
\email{Aurelien.Barrau@cern.ch}
\affiliation{
Laboratoire de Physique Subatomique et de Cosmologie, Universit\'e Grenoble-Alpes, CNRS-IN2P3\\
53, Avenue des Martyrs, 38026 Grenoble cedex, France\\
}%

\author{Pierre Jamet}
\email{pierre.jamet@vivaldi.net}
\affiliation{
Laboratoire de Physique Subatomique et de Cosmologie, Universit\'e Grenoble-Alpes, CNRS-IN2P3\\
53, Avenue des Martyrs, 38026 Grenoble cedex, France\\
}%

\author{Killian Martineau}
\email{martineau@lpsc.in2p3.fr}
\affiliation{
Laboratoire de Physique Subatomique et de Cosmologie, Universit\'e Grenoble-Alpes, CNRS-IN2P3\\
53, Avenue des Martyrs, 38026 Grenoble cedex, France\\
}%

\author{Flora Moulin}
\email{flora.moulin@lpsc.in2p3.fr}
\affiliation{
Laboratoire de Physique Subatomique et de Cosmologie, Universit\'e Grenoble-Alpes, CNRS-IN2P3\\
53, Avenue des Martyrs, 38026 Grenoble cedex, France\\
}

\begin{abstract}
This article is devoted to the study of scalar perturbations in loop quantum cosmology. It aims at clarifying the situation with respect to the way initial conditions are set and to the specific choice of an inflaton potential. Several monomial potentials are studied. Both the dressed metric and deformed algebra approaches are considered. We show that the calculation of the ultraviolet part of the spectrum, which is the physically relevant region for most background trajectories, is reliable, whereas the infrared and intermediate parts do depend on some specific choices that are made explicit. 
\end{abstract}

\maketitle

\section{Introduction}

The calculation of primordial cosmological power spectra is an important way to connect speculative theories of quantum gravity with observations (see \cite{Barrau:2017tcd} for a recent review). Among those theories, loop quantum gravity (LQG) (see, {\it e.g.}, \cite{Rovelli:2014ssa}) has now reached the point where explicit calculations can be performed. At this stage, it remains, however extremely difficult to derive rigorous cosmological predictions from the full theory. But, in the specific case of loop quantum cosmology (LQC), which can be viewed as the quantization of symmetry reduced general relativity using techniques from LQG (see, {\it e.g.}, \cite{lqc9,Ashtekar:2015dja}), quite a lot of results have already been obtained, beginning with the replacement of the usual Big Bang by a Big Bounce. Recently, important improvements were proposed, {\it e.g.} in group field theory  \cite{Gerhardt:2018byq,Oriti:2016acw,Oriti:2016ueo}, in quantum reduced loop gravity \cite{Alesci:2016xqa,Alesci:2016gub,Alesci:2015nja,Alesci:2013xd}, in refined coherent state approaches \cite{Dapor:2018sgb}, in diffeomorphism invariance derivation \cite{Engle:2018zbe} or in analogies with a Kasner transition \cite{Wilson-Ewing:2017vju}, to cite only a few. \\

Together with hybrid quantization \cite{Fernandez-Mendez:2013jqa,Blas:2014naa}, two main approaches have been developed in this framework to study inhomogeneities: the dressed metric \cite{Agullo1,Agullo2,Agullo3} and the deformed algebra \cite{tom1,tom2,eucl2,eucl3}. The first deals with quantum fields on a quantum background while the second puts the emphasis on the consistency and covariance of the effective theory. This led to clear predictions about the power spectra \cite{Agullo:2015tca,Bolliet:2015bka,lcbg,Schander:2015eja,Martineau:2017tdx,Bolliet:2015raa}. Other complementary paths were also considered to investigate perturbations \cite{ed,Gielen:2017eco,Agullo:2018wbf,Wilson-Ewing:2016yan,ElizagaNavascues:2017avq}. \\

Many works were devoted to tensor perturbations that are easier to handle both for gauge and for anomaly issues. Scalar modes are, however more important from the observational viewpoint (see, {\it e.g.}, \cite{Agullo:2015tca,Bonga:2015kaa,Bonga:2015xna,Agullo:2016hap,Ashtekar:2016pqn} for recent works in LQC). This article focuses on scalar spectra and aims at clarifying how previous LQC results obtained for a simple massive scalar field can be generalized to other monomial potentials and to which extent the spectrum is sensitive to initial conditions ({\it i.e.} to a vacuum choice) for perturbations. It is essentially impossible to derive fully generic results, so we explicitly investigate different solutions and show the associated numerical computations so that they can be accounted for in future studies.

\section{Generic framework}

We consider here a spatially flat and isotropic FLRW spacetime filled with a minimally coupled scalar field with a monomial potential. We neglect backreaction and trans-Planckian effects. \\

We first come back to the study developed by some of the authors of this article in \cite{Schander:2015eja}. As in this work, we adopt here a causal viewpoint and put the initial conditions, both for the background and the perturbations, as far as possible in the contracting phase preceding the bounce.  \\

The basic ingredients are the following. The Friedmann equation, modified by holonomy corrections, reads as

\begin{equation}
\label{eq:fried_new}
H^2 = \frac{\kappa}{3}\rho\left(1-\frac{\rho}{\rho_c}\right),
\end{equation}

where $\rho_c$ is the critical density (expected to be of the order of the Planck density), and $H = \dot{a}/a$ is the Hubble parameter. The Klein-Gordon equation for  the background is given by 

\begin{equation}
\label{eq:kg_new}
\ddot{\varphi} = -3H\dot{\varphi} - \partial_\varphi V(\varphi),
\end{equation}

where $\varphi$ is here used for $\bar{\varphi}$, the average scalar field. The differential system for the background can be summarized as (we choose the convention  $a(t_{init})=1$):

\begin{align}
\dot{\varphi}(t) &= \frac{\partial \varphi}{\partial t},\\
\ddot{\varphi}(t) &= -3H(t)\dot{\varphi}(t) - \partial_\varphi V(\varphi(t)),\\
\dot{H}(t) &= -\frac{\kappa}{2}\dot{\varphi}^2(t)\left(1-2\frac{\dot{\varphi}^2(t)/2 + V(\varphi(t))}{\rho_c}\right),\\
\dot{a}(t) &= H(t)a(t).
\end{align}

Perturbations are described in the Fourier space by the gauge invariant Mukhanov-Sasaki equation:

\begin{equation}
\label{eq:m_s_fourier}
v^{\prime\prime}_k + \left(k^2 - \frac{z^{\prime\prime}}{z}\right)v_k= 0,
\end{equation}

where $z = \frac{a\dot{\varphi}}{H}$, and the derivation is with respect to the conformal time $d\eta = \frac{1}{a}d t$. One can easily show that 

\begin{equation}
\frac{\ddot{z}}{z} = \frac{\dddot{\varphi}}{\dot{\varphi}} + \frac{\ddot{\varphi}}{\dot{\varphi}}\left(2H - 2\frac{\dot{H}}{H}\right) + H^2 - \dot{H} + 2\left(\frac{\dot{H}}{H}\right)^2 - \frac{\ddot{H}}{H}.
\end{equation}

Introducing
\begin{equation}
\Omega = 1-2\frac{\rho}{\rho_c},
\end{equation}

and using
\begin{equation}
\label{eq:rho_new}
\rho = \frac{1}{2}\dot{\varphi}^2 + V(\varphi),
\end{equation}

this leads to the final expression:

\begin{widetext}
\begin{equation}
\frac{z^{\prime\prime}}{z} = a^2\left(-\partial^2_\varphi V(\varphi) + 2H^2 - 2\kappa\Omega\frac{\dot{\varphi}\partial_\varphi V(\varphi)}{H} - \frac{7}{2}\kappa\Omega\dot{\varphi}^2 + \frac{3\kappa}{\rho_c}\dot{\varphi}^4 + \kappa^2\Omega^2\frac{\dot{\varphi}^4}{2H^2}\right).
\end{equation}
\end{widetext}

This is the intricate effective potential that has to be dealt with. In the next two sections, we study perturbations as described by the dressed metric approach \cite{Agullo1,Agullo2,Agullo3}, which is very close to the hybrid quantization one as far as phenomenology is concerned \cite{Wilson-Ewing:2016yan}. Interestingly, at the effective level, the equation of motion (\ref{eq:m_s_fourier}) is formally the same than in general relativity, even though the value of $z''/z$ is of course heavily modified. We then switch to the deformed algebra approach were an effective change of signature shows up.  

\section{Quadratic potential}

\begin{figure}[!h]
\begin{center}
\includegraphics[scale=0.35]{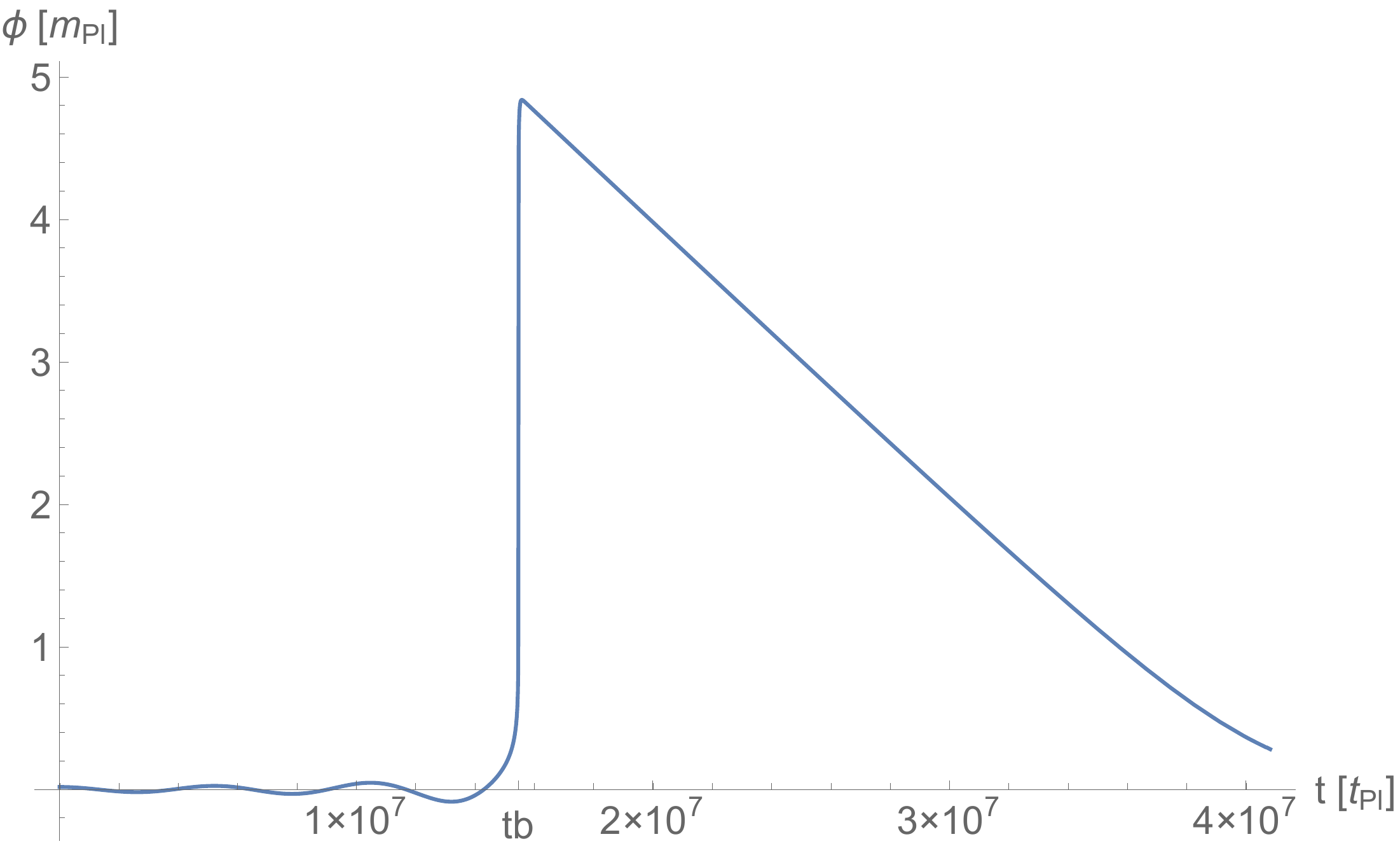}
\caption {Temporal evolution of the scalar field, for a mass $m=1.2\times10^{-6}$.} 
\label{field1}
\end{center}
\end{figure}

The resulting typical evolution of the scalar field is shown in Fig \ref{field1}: pseudo-oscillations are followed by the bounce and by an inflationary stage. The details obviously depend on the phase of the field during the contracting period but, as shown in \cite{bl,Linsefors:2014tna,Martineau:2017sti}, what is displayed in  Fig \ref{field1} is a quite generic behavior. The probability to have, for example, a phase of deflation is much smaller. All numbers are given in Planck units.

The way to choose initial conditions for the perturbations is more subtle. The usual Minkowski solution
\begin{equation}
v_k(\eta) = \frac{1}{\sqrt{2k}}e^{-ik\eta},
\end{equation}
is approached in the so-called Bunch-Davies vacuum. The main requirement to set the vacuum is that the effective potential is negligible so that the equation of motion becomes nearly the one of an harmonic oscillator. In addition, if the causal evolution of the Universe during the bounce is taken seriously and if the word ``initial" is taken literally, it makes sense to put initial conditions far away before the bounce, this later constituting in addition the most ``quantum'' and less controlled moment in the whole cosmic history (see {\it e.g.} \cite{Barrau:2016nwy} for a discussion). As it will become clear later, this requirement is actually in tension with the first one (which should be considered as the mandatory one).\\

\begin{figure}[!h]
\begin{center}
\includegraphics[scale=0.35]{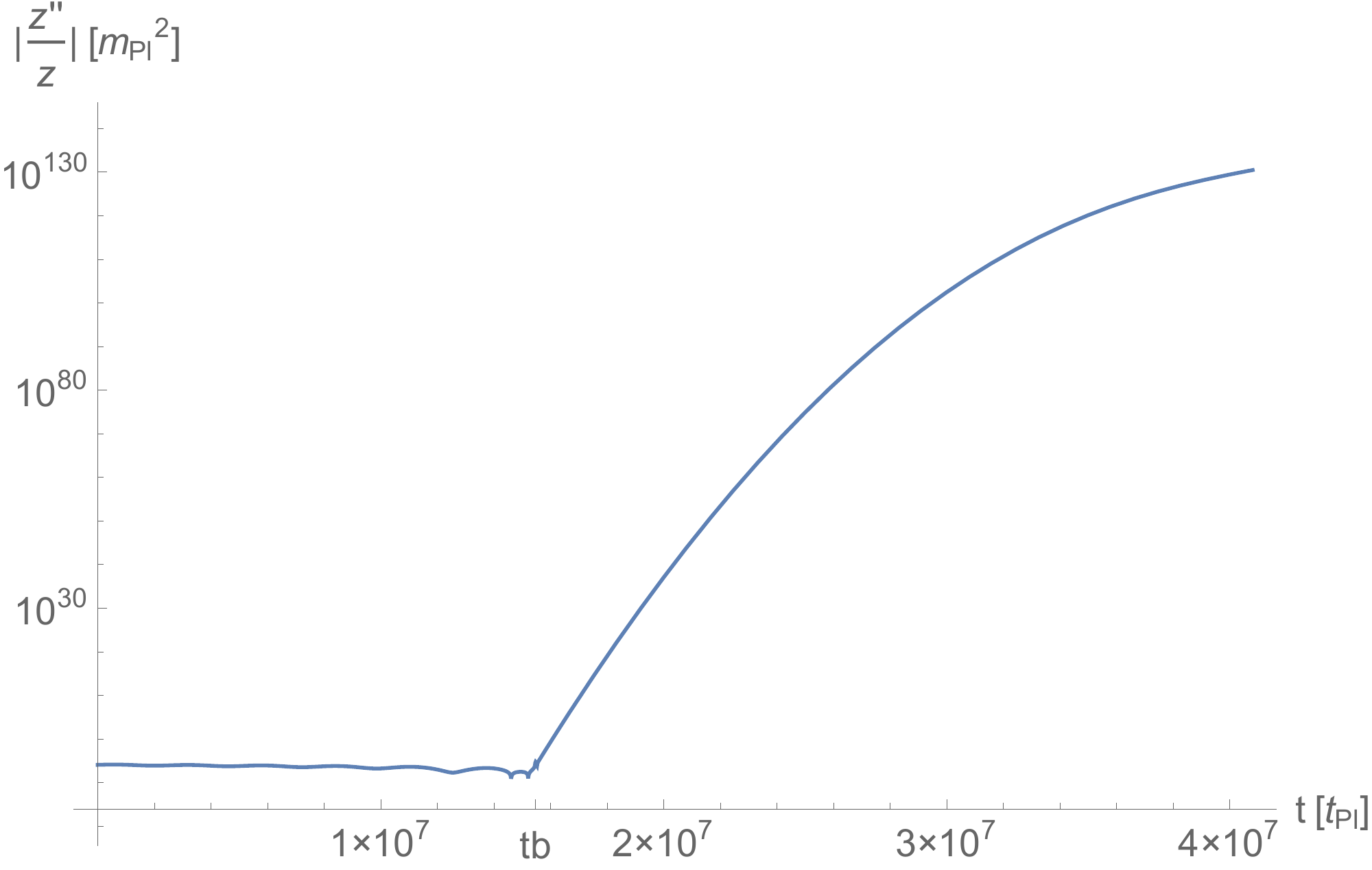}
\caption{Temporal evolution of the absolute value of effective potential $\frac{z^{\prime\prime}}{z}$ over the full integration interval.} 
\label{effpotquad2}
\end{center}
\end{figure}

The evolution of the absolute value of the effective potential $\frac{z^{\prime\prime}}{z}$ is shown in Fig. \ref{effpotquad2} during the full integration time interval. It should be noticed that it increases both in the past and in the future of the bounce (which is located around $t=1.5\times 10^{7}$ on the plot). This raises an issue which is fundamental for bouncing models and should be taken into account with care, as studied later in this article.\\

\begin{figure}[!h]
\begin{center}
\includegraphics[scale=0.42]{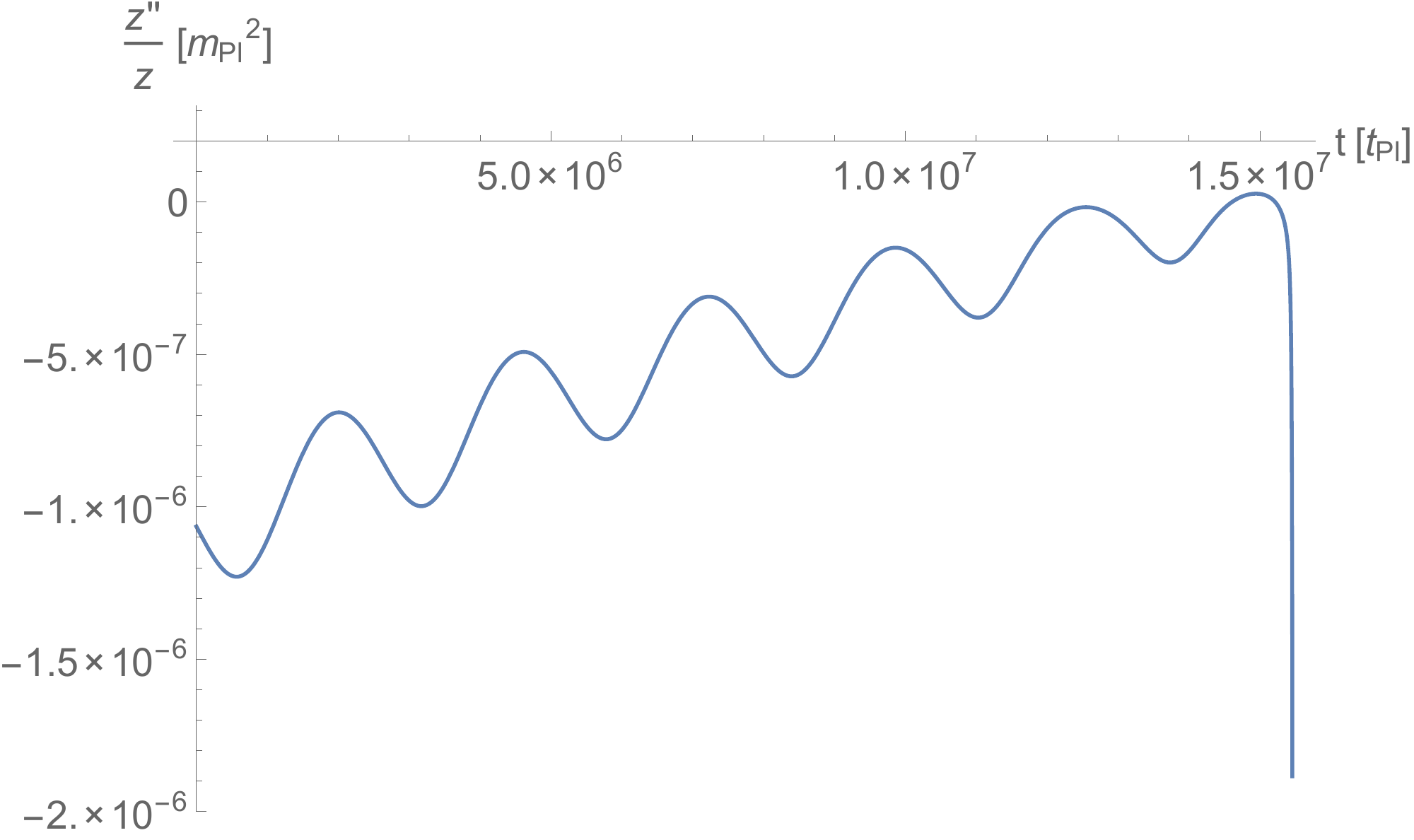}
\caption{Temporal evolution of the effective potential $\frac{z^{\prime\prime}}{z}$ between the beginning of the integration and just before the bounce.} 
\label{effpotquad1}
\end{center}
\end{figure}

Figure \ref{effpotquad1} shows the effective potential between the beginning of the integration interval and the bounce. The shape is highly complex and very different from what happens either in standard cosmology or in LQC for tensor modes. In the standard cosmological model it vanishes when going backward in time, deep into the de Sitter inflationary phase. This is also true for bouncing models when going far away in the past of the contracting phase, but only for tensor modes. In the considered case, due to the large (negative) value taken by the potential in the remote past it is impossible to put stable initial conditions very far from the bounce. Strictly speaking, it might make sense to set initial conditions in this way but the interesting selection criterion associated with the Bunch-Davies vacuum would be lost. If one wants to remain in a framework where a Bunch-Davies like initial state -- which is at least justified to compare with other results -- is used, there are two moments which can be chosen such that $\frac{z^{\prime\prime}}{z}$ vanishes. However, those points are not far from the bounce and the fact that ``initial'' conditions have to be set at very specific moments is something that deserves to be better understood in the future and should be, at this stage, considered as a weakness (at least at the heuristic level) of those models.\\

\begin{figure}[!h]
\begin{center}
\includegraphics[scale=0.38]{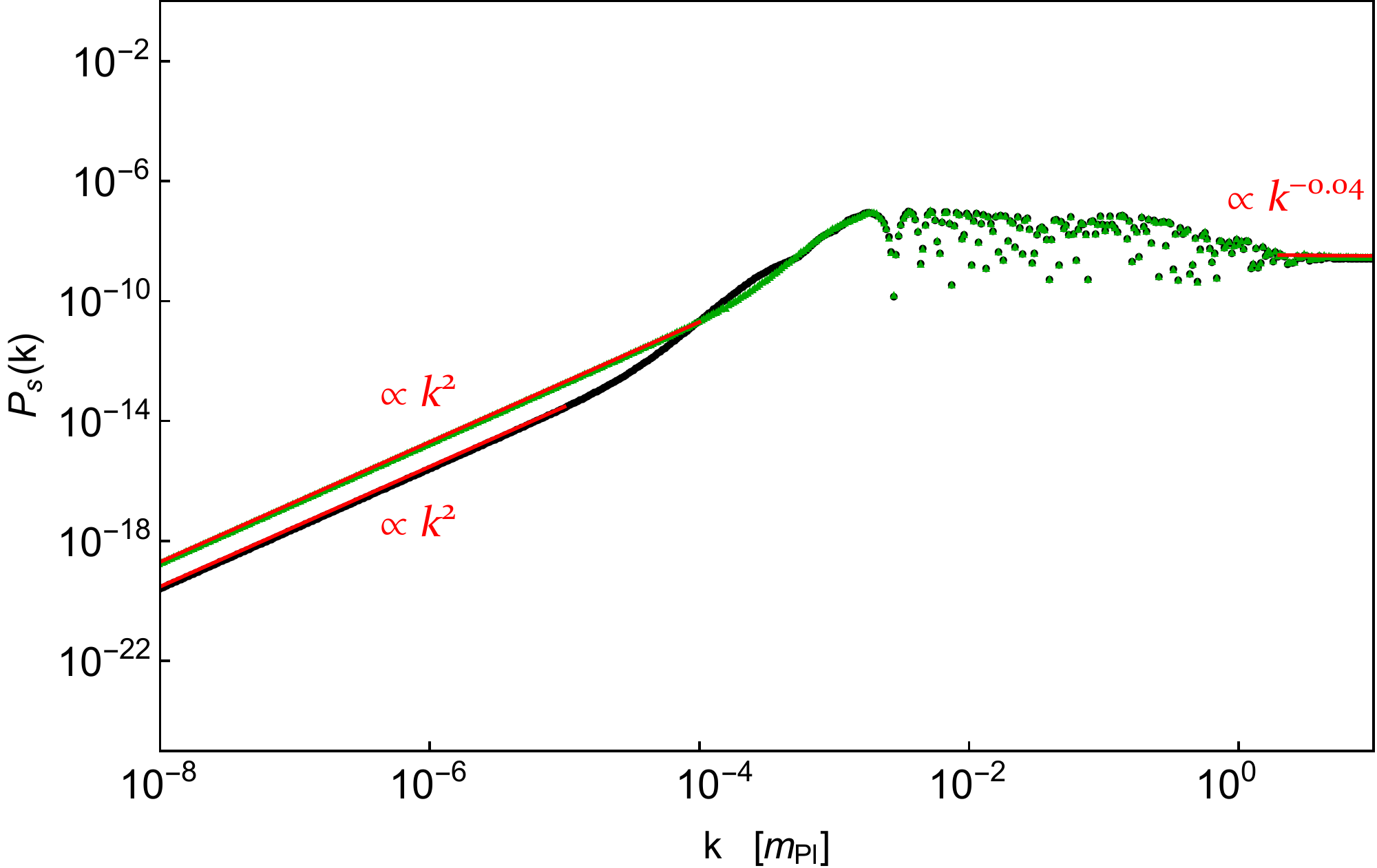}\
\caption{Primordial scalar power spectra, as a function of the comoving wave number, for a quadratic potential and initial conditions put whether at the first zero of $\frac{z^{\prime\prime}}{z}$, \textit{i.e} $t_{i}=1.46\times10^{7}$ (lower plot in the IR, black disks), or at the last zero of $\frac{z^{\prime\prime}}{z}$, \textit{i.e}  $t_{i}=1.52\times10^{7}$ (upper plot in the IR, green triangles).}
\label{spquad}
\end{center}
\end{figure}

As a first step in a better understanding of the situation, we present in Fig. \ref{spquad} the primordial power spectra resulting from a full simulation of the evolution of perturbations with initial conditions set both at the first zero,  \textit{i.e} at $t_{i}=1.46\times10^{7}$, corresponding to the earliest time in cosmic history, and at the second zero of the effective potential at $t_{i}=1.52\times10^{7}$.\\

First, it should be emphasized that the ultraviolet (UV) part of the spectrum is the same for both ways of putting initial conditions and is compatible with observations, that is nearly scale invariant with a very slight tilt due to the slow roll of the field during the inflationary stage. This is particularly important as the UV part of the spectrum is most probably the one which is experimentally probed. This last fact entirely depends on the number of e-folds of inflation: the conversion of the comoving wave number into a physical wave number requires the knowledge of the expansion factor of the Universe. Except if the background initial conditions are hyper-fine-tuned, inflation lasts long enough \cite{bl,Linsefors:2014tna,Martineau:2017sti} so that the observational cosmological microwave background (CMB) window clearly falls in the UV part of the spectrum. In principle, this would require a specific trans-Planckian treatment (see \cite{Espinoza-Garcia:2017qjl,Martineau:2017tdx} for first attempts in this direction) which is not the topic of this study and which is anyway partially accounted for in the dressed metric approach. The oscillations in the intermediate part of the spectra -- due to quasi-bound states in the effective Shr\"odinger equation -- are basically the same in both cases, together with the deep infrared (IR) part (throughout all the article we call ``infrared" the rising part of the spectrum and ``ultraviolet" the scale-invariant one).  However, some differences do remain in the junction between the IR and the oscillatory regimes. We have checked that they are not due to numerical issues. Although this is not of high phenomenological significance, this shows that the way initial conditions are set, even around a vanishing effective potential, can influence the resulting power spectrum. \\

\begin{figure}[!h]
\begin{center}
\includegraphics[scale=0.38]{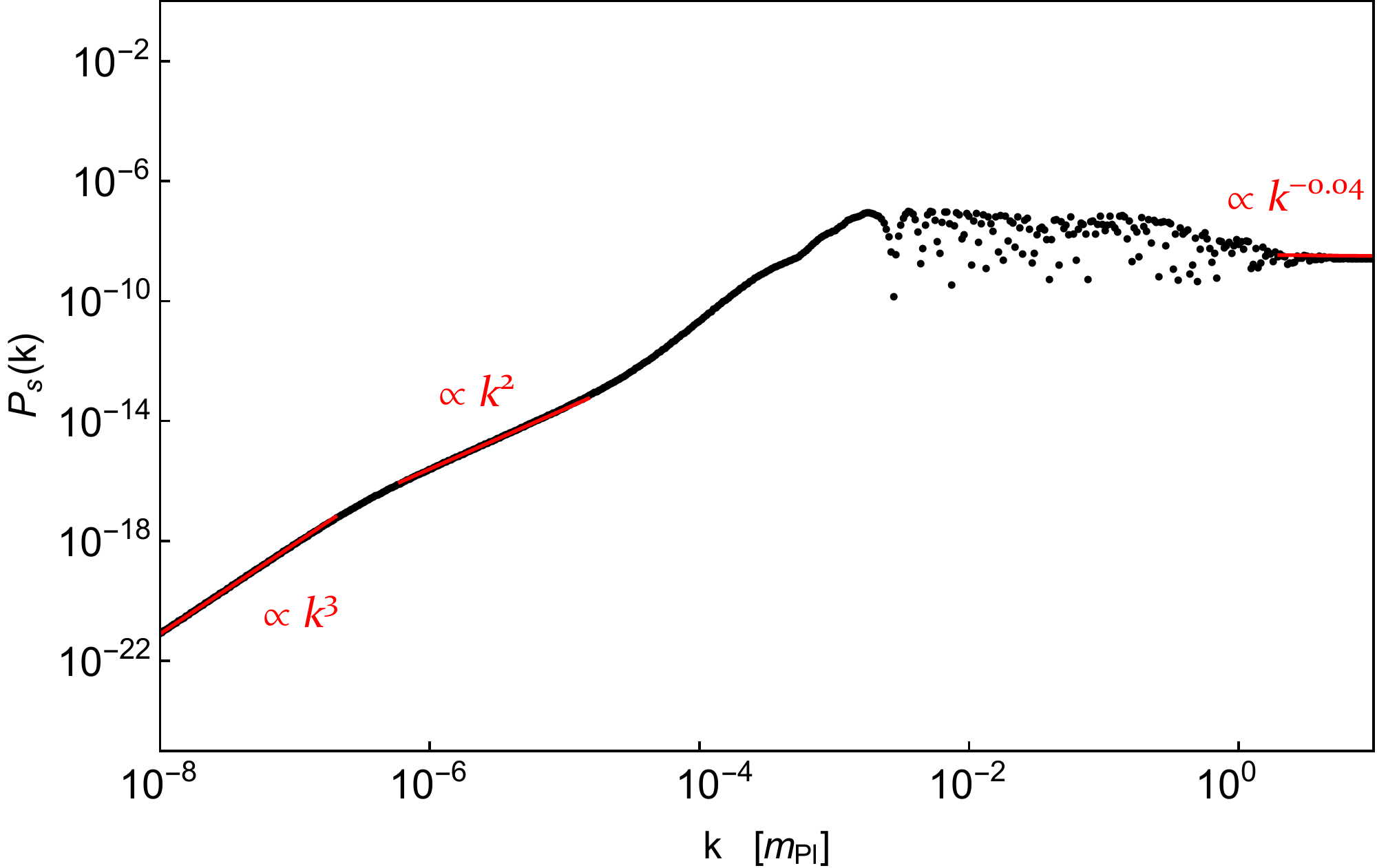}
\caption{Primordial scalar power spectrum, as a function of the comoving wave number, for a quadratic potential and initial conditions set  $0.6 t_{\text{Pl}}$ before the Bunch-Davies vacuum at $t_{i}=1.46\times10^{7}$.}
\label{spquad3}
\end{center}
\end{figure}

We have also checked that when moving slowly away from the exact point were $\frac{z^{\prime\prime}}{z}=0$, the spectrum slowly changes. This is obviously expected but the details of the changes are very hard to guess as the effective potential is very complicated. Basically, the spectrum evolves from a full $k^2$ to a full $k^3$ behavior in the IR. Figure \ref{spquad3} presents an intermediate case, and this should be taken into account when interpreting results given in \cite{Schander:2015eja}.\\

\begin{figure}[!h]
\begin{center}
\includegraphics[scale=0.38]{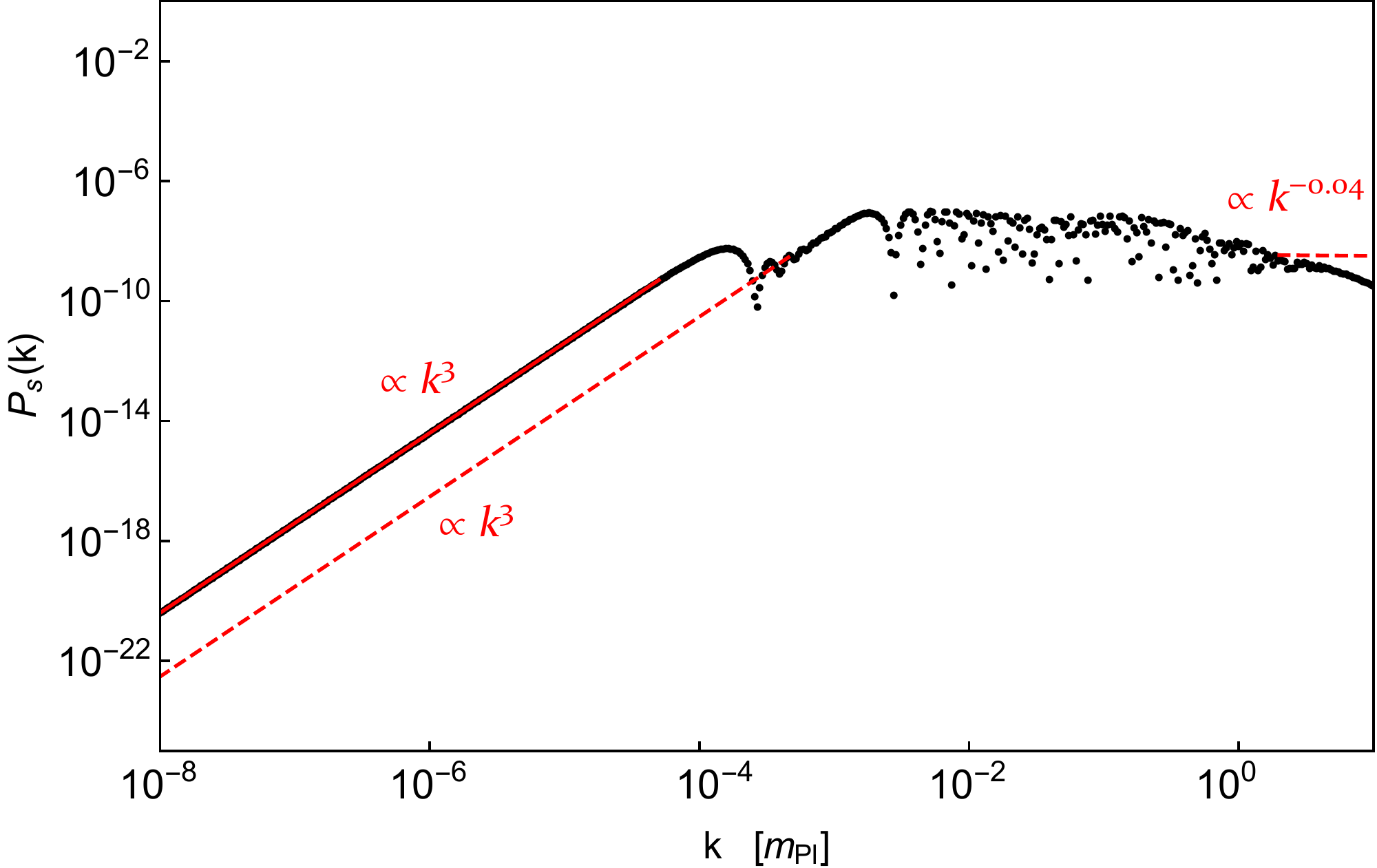}
\caption{Primordial scalar power spectrum, as a function of the comoving wave number, for a quadratic potential and initial conditions at local extrema far away from the bounce, at $t_{i}=2.00\times10^{6}$. From the dot line to the plain line, one goes deeper in the remote past.} 
\label{spquad4}
\end{center}
\end{figure}

Finally, in Fig. \ref{spquad4}, the spectrum is plotted for initial conditions set at a local extremum further away from the bounce, at $t_{i}=2.00\times10^{6}$. The plain line corresponds to a point deeper in the past than the dotted line.\\

This shows that although the global shape of the spectrum is under control -- especially in the region of phenomenological significance -- the detailed structure is quite sensitive to the way initial conditions are set. In models where the effective potential does not vanish in the remote past, this raises nontrivial issues. This means by no way that those approaches are inconsistent but that some uncertainties associated with the loss of a strong selection criterion on initial conditions have to be included in the analysis.

\section{Generalized potentials}

It is important to investigate whether the scalar spectra obtained hold for other inflaton potential shapes (not to be confused with the effective potential felt by pertubations), beyond the massive scalar field which is not favored by data \cite{Ade:2015lrj}. The case of plateaulike potentials is very specific in bouncing models (see \cite{Martineau:2017sti}), so we restrain ourselves to confining monomial potentials of the form:

\begin{equation}
V(\varphi) = \frac{1}{n}\lambda_n{\varphi}^n.
\end{equation}

\begin{figure}[!h]
\begin{center}
\includegraphics[scale=0.38]{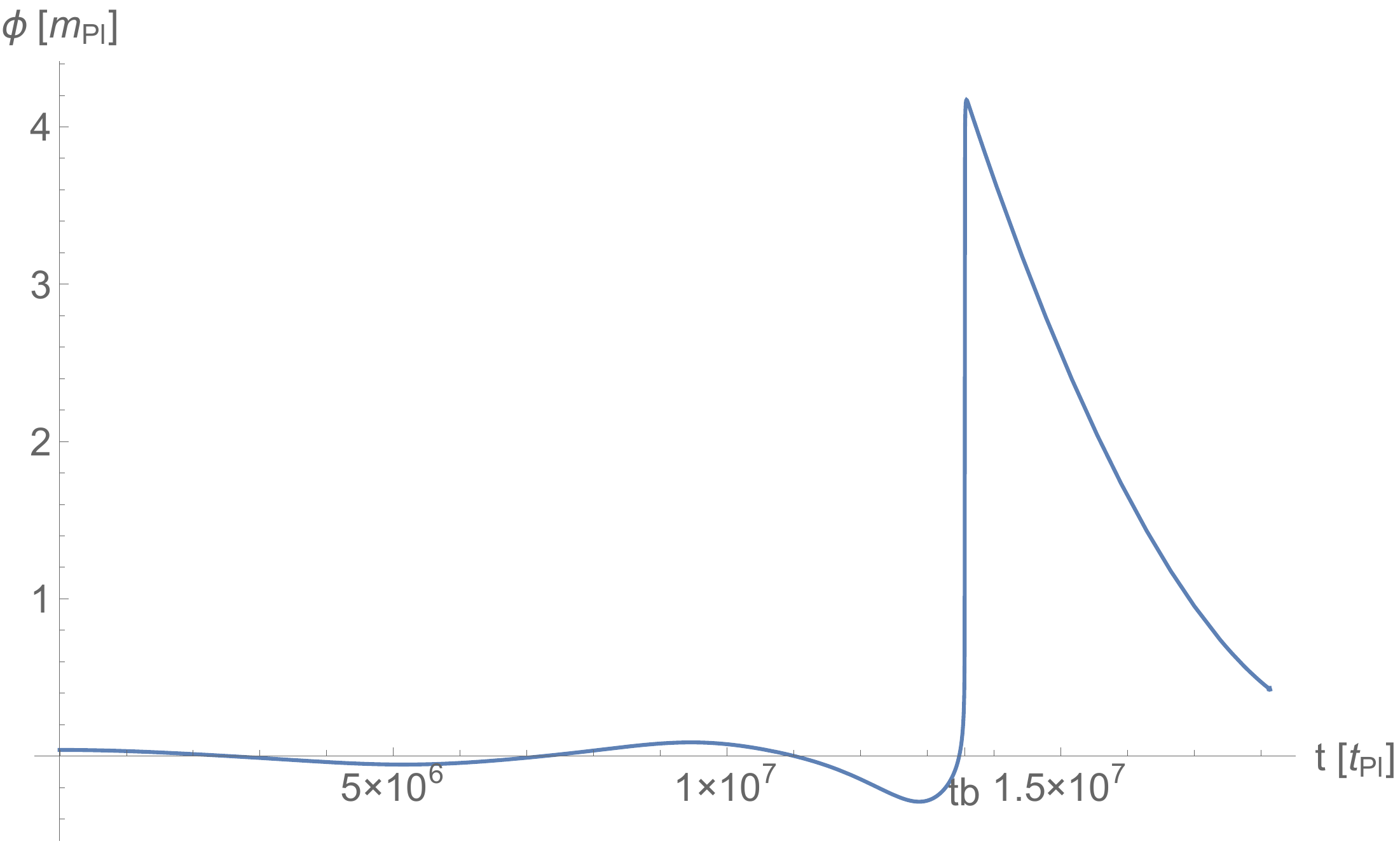}
\caption{Evolution of the scalar field for $n=3$.} 
\label{phin3}
\end{center}
\end{figure}

No general analytical solution in the deep contracting phase can be found anymore but it is still possible to set initial conditions for the background as done previously. The evolution of the scalar field is qualitatively weakly depending on $n$. As an example, we show the result for $n=3$ in Fig. \ref{phin3}.\\

\begin{figure}[!h]
\begin{center}
\includegraphics[scale=0.38]{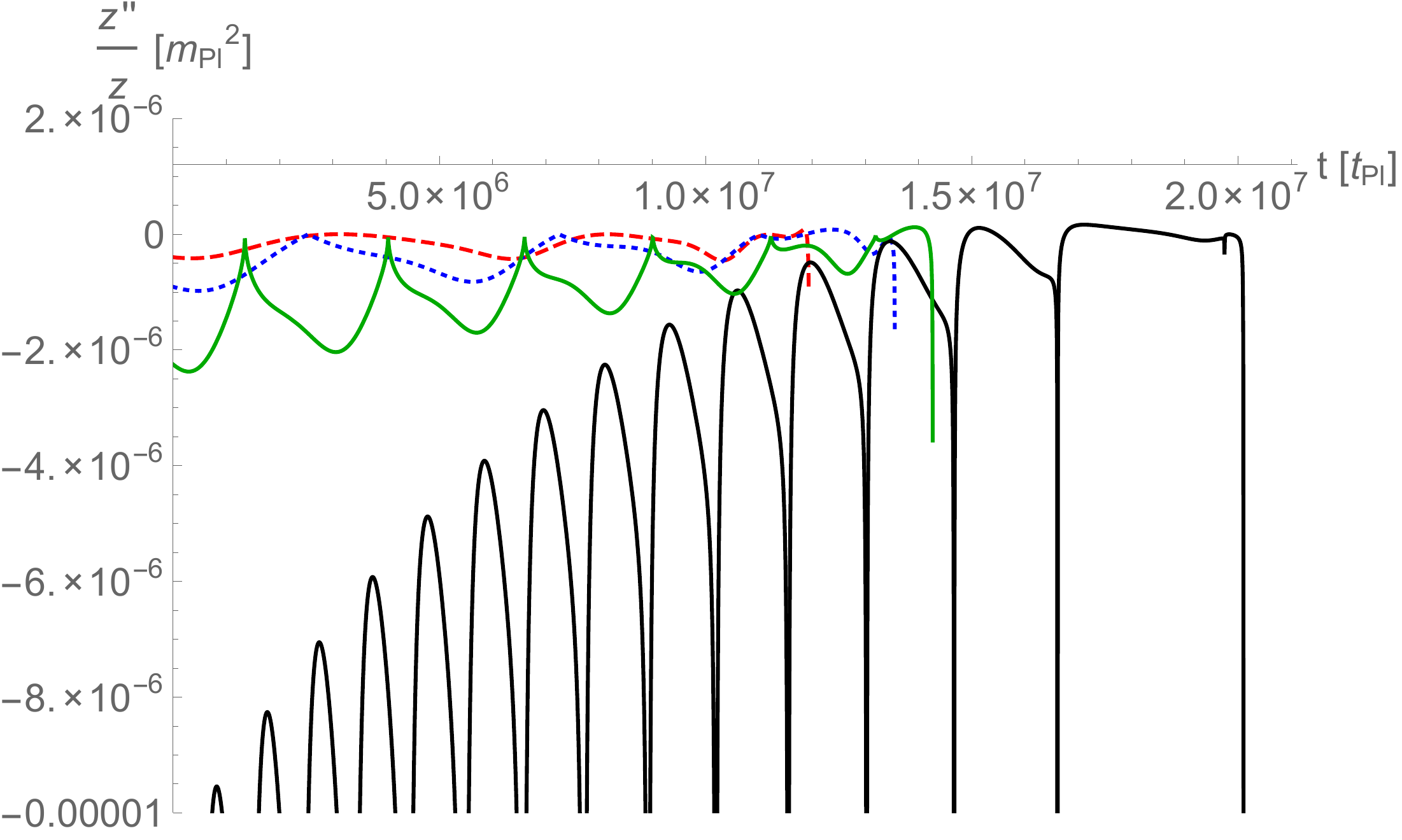}
\caption{Evolution of $\frac{z^{\prime\prime}}{z}$ in the contracting universe up to the respective bounces for $V(\varphi) = \frac{1}{n}\lambda_n{\varphi}^n$ and $n=3$ (blue dotted line), $n=4$ (red dashed line), $n=4/3$ (black solid line) and $n=5/2$ (green solid line).}
\label{potgenall}
\end{center}
\end{figure}

The situation is more complicated when one considers the details of the effective potential. Figure \ref{potgenall} shows the evolution of $z^{\prime\prime}/z$ up to the respective bounces for $n=3,4,4/3,5/2$.

Clearly the shape of the behavior of the effective potential depends on the value of $n$. The number of points were the potential identically vanishes is finite in each case, leading to a finite number of ways to set a rigorous instantaneous Bunch-Davis vacuum. In all cases there is also an infinite number of local minima that can be used as approximate vacua, depending on the range of wave numbers relevant for the considered study. We insist once more that the details of the spectrum {\it do} depend on this choice.\\

\begin{figure}[!h]
\begin{center}
\includegraphics[scale=0.38]{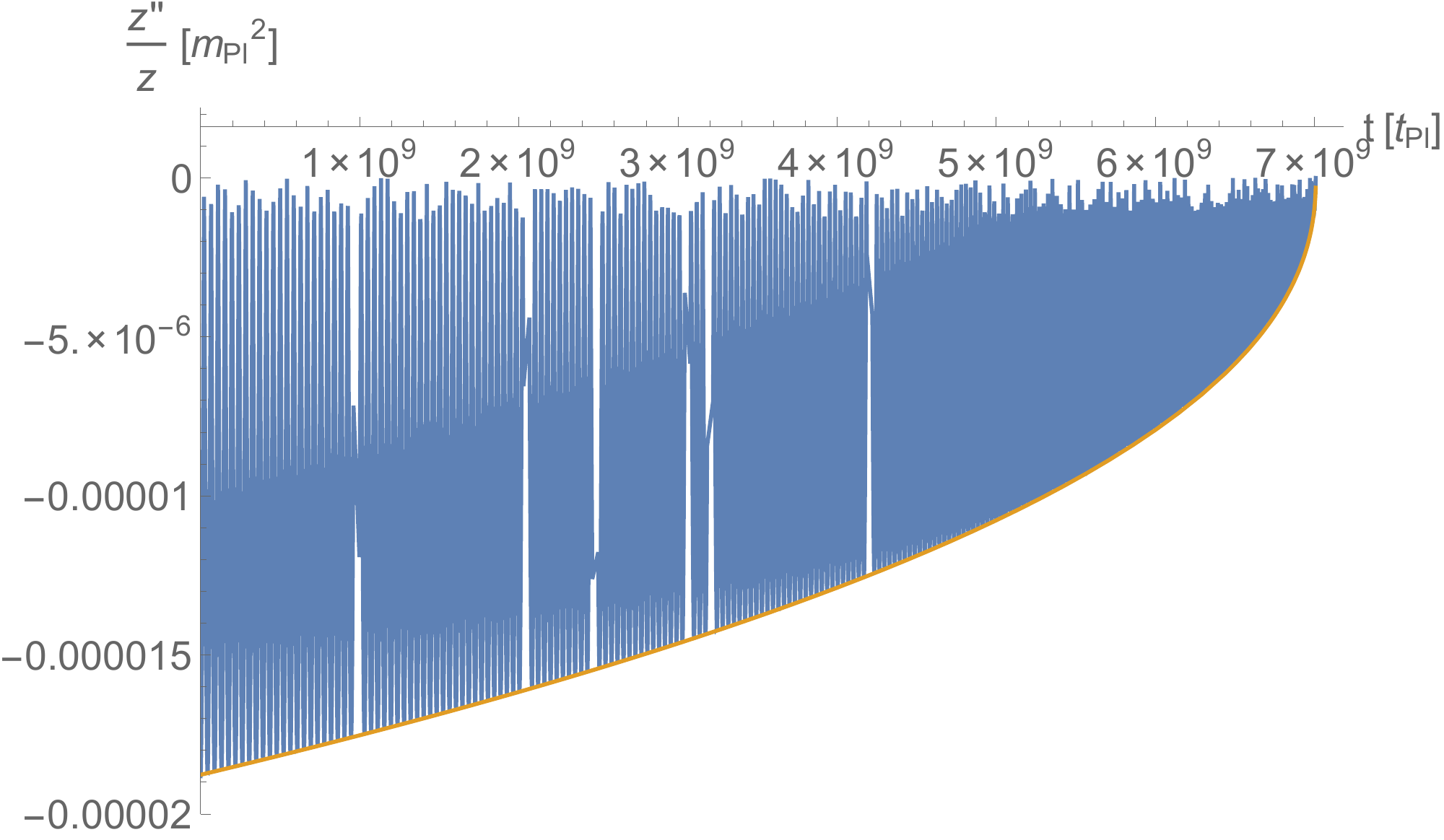}
\caption{Evolution of $\frac{z^{\prime\prime}}{z}$, together with its envelop, in the contracting universe, up to the bounce on a wide time interval.} 
\label{envelop}
\end{center}
\end{figure}

Figure \ref{envelop} shows the envelope of the effective potential for $n=3$ when going deeper into the past. It can be empirically fitted by a power law $(t-t_b)^{0.45}$. The oscillations themselves get quite chaotic, reflecting the nonlinearity of the equations. The situation is very different from what happens for the effective potential of tensor modes. It might be that, from the bounce, time flows in two opposite directions. Then it would make sense to put initial conditions at the bounce, as in \cite{Agullo1,Agullo2,Agullo3}. If, however, the evolution remains globally causal with a unique time direction, the questions raised here cannot be ignored.\\

\begin{figure}[!h]
\begin{center}
\includegraphics[scale=0.38]{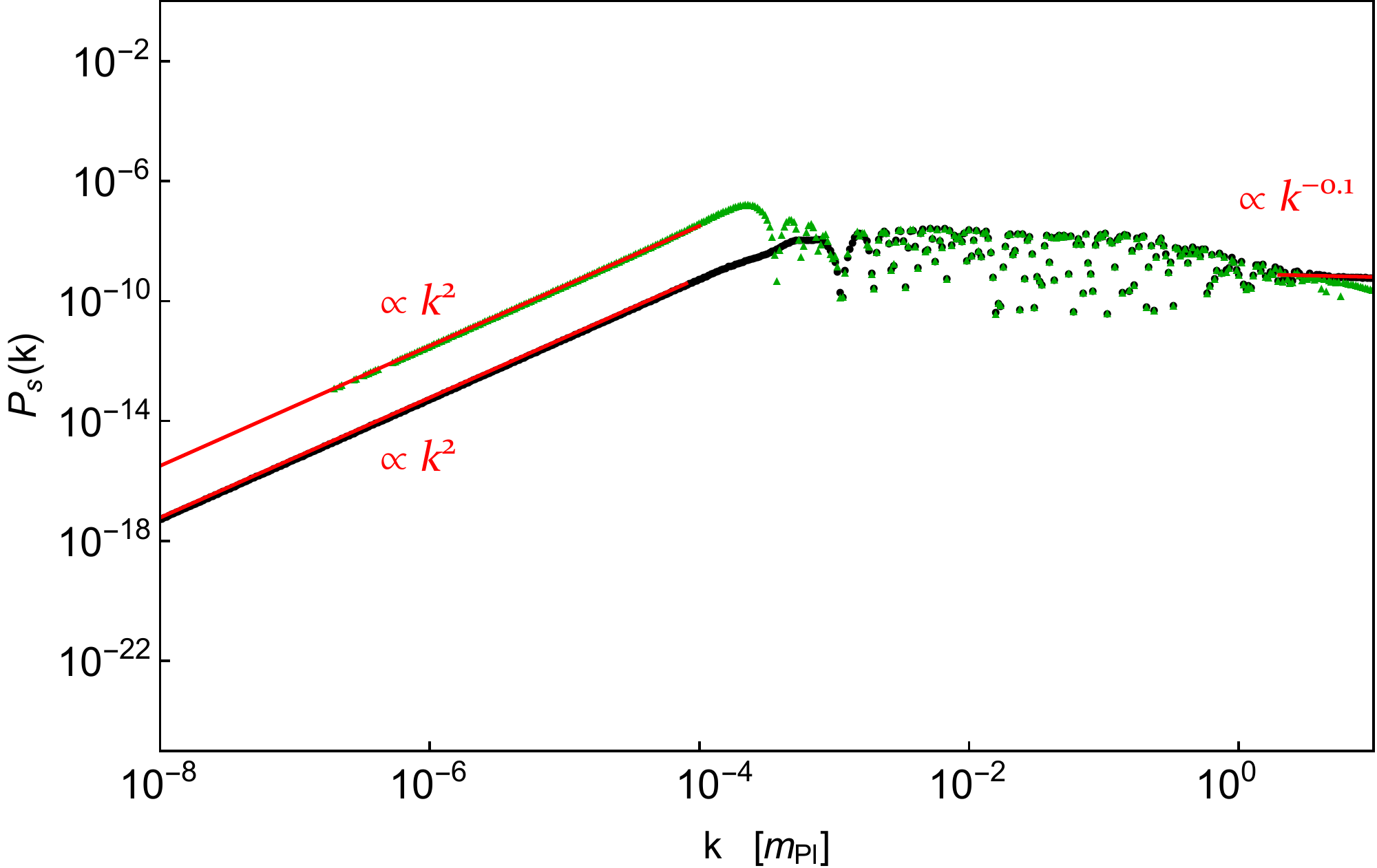}
\caption{Spectrum for $n = \frac{4}{3}$ with initial conditions whether close to the bounce, at $t_{i}=1.87\times10^{7} t_{\text{Pl}}$ (lower plot in the IR, black disks), or far from it at $t_{i}=1.49\times10^{7}$ (upper plot in the IR, green triangles).}
\label{specgen1}
\end{center}
\end{figure}

\begin{figure}[!h]
\begin{center}
\includegraphics[scale=0.38]{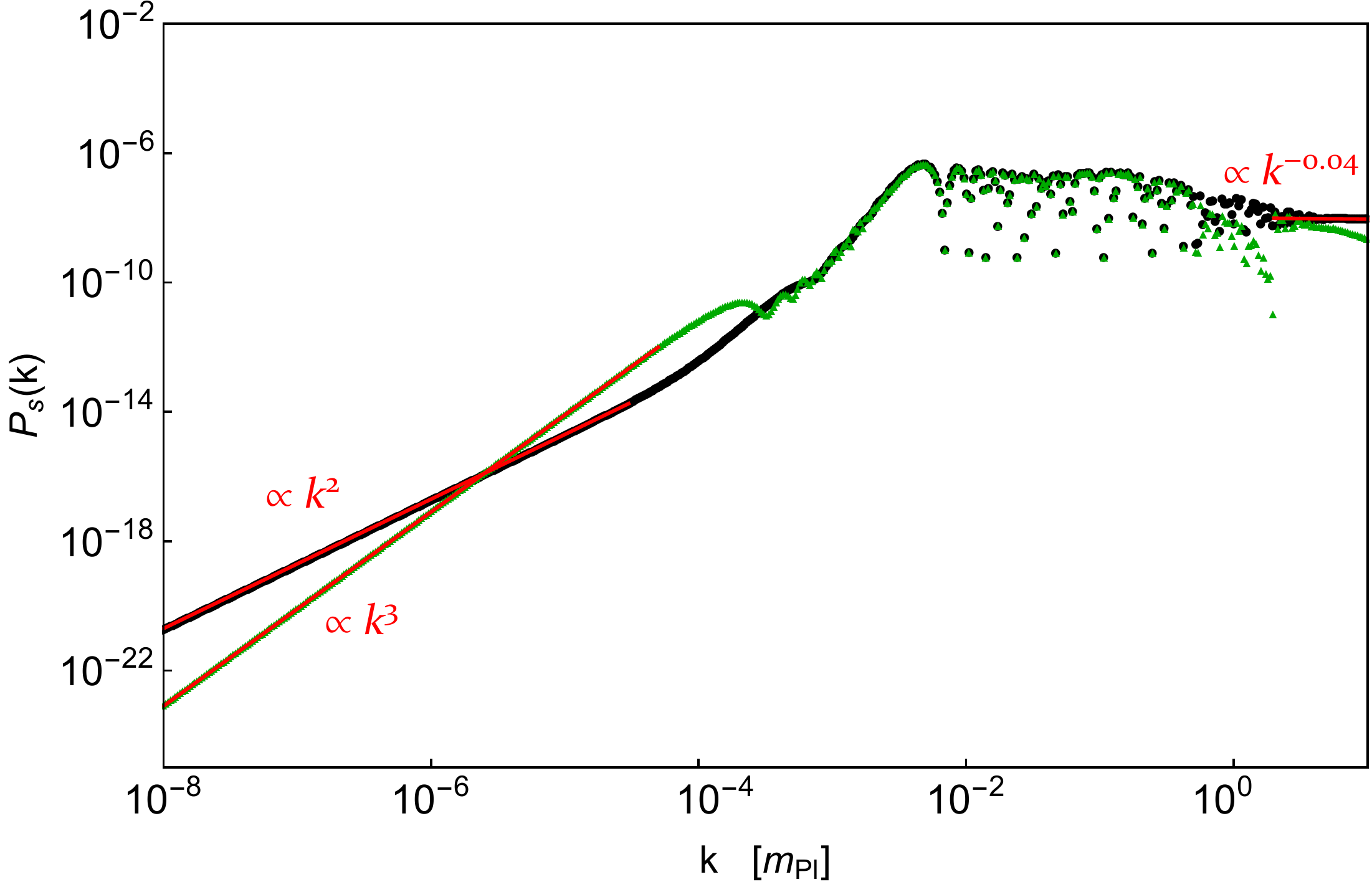}
\caption{Spectrum for $n = \frac{5}{2}$ with initial conditions whether close to the bounce, at $t_{i}=1.35\times10^{7}$ ($k^{2}$ behavior in the IR, black disks), or far from it at $t_{i}=1.35\times10^{6} t_{\text{Pl}}$ ($k^{3}$ behavior in the IR, green triangles).}
\label{specgen2}
\end{center}
\end{figure}

As ``extreme'' examples, we show in Fig. \ref{specgen1} (resp. Fig. \ref{specgen2}) the scalar spectra for $n = 4/3$ (resp. $n = 5/2$) with initial conditions set close to the bounce and in the deep past. This reinforces the previous conclusions: the ``small scales'' part of the spectra is nearly scale invariant in all cases (although small differences do exist), making the results compatible with observation for the vast majority of the parameter space which leads to an inflationary stage so long that the observable part falls in the deep UV range. However, the IR part and some of the oscillations can sensitive to the details of the inflaton potential shape and to the way initial conditions are set. 

\section{Deformed algebra}

Another approach to LQC, the so-called deformed algebra, relies on a different view of the situation \cite{Bojowald:2011aa,eucl3,tom1,tom2,eucl2,Cailleteau:2013kqa}. In this case, the emphasis in put on the consistency of the effective theory. The Poisson brackets are calculated between (holonomy) quantum corrected constraints. Anomalies do appear in general. To ensure covariance, counter-terms with a vanishing classical limit are added to the constraints, so that the system remains ``first class" in the Dirac sense. The resulting algebra (including the matter content) is closed and reads as

\begin{eqnarray}
\left\{ D[N^a_1], [N^a_2] \right\} &=& 0, \\
\left\{ H[N],D[N^a] \right\} &=& - H[\delta N^a \partial_a \delta N], \\
\left\{ H[N_1],H[N_2] \right\} &=&  D \left[  \Omega  \frac{\bar{N}}{\bar{p}} \partial^a(\delta N_2 - \delta N_1)\right],  
\label{HtotHtot}
\end{eqnarray}

where $D[N^i]$ is the full diffeomorphism constraint and $H[N]$ is the full scalar constraint. The important feature it the $\Omega = (1-2\rho / \rho_c)$ term in the last Poisson bracket. It becomes negative close to the bounce and leads to an effective change of signature. The Mukhanov equation of motion in Fourier space reads, in this framework, as

\begin{equation}
\ddot{\mathcal{R}}_k - \left( 3H + 2 m^2 \frac{\bar{\varphi}}{\dot{\bar{\varphi}}} + 
2 \frac{\dot{H}}{H} \right) \dot{\mathcal{R}}_k + \Omega\frac{k^2}{a^2} \mathcal{R}_k = 0,
\label{eomR}
\end{equation}

with 
\begin{equation}
\mathcal{R} := \frac{v}{z},
\end{equation}
where $v$ is the gauge-invariant perturbation  and $z$ is the background variable. Phenomenologically, the main consequence of this model, if a causal view is chosen and a massive scalar field is assumed to fill the Universe, is an exponential growth of the spectrum in the UV \cite{lcbg,Schander:2015eja}. It is obviously not compatible with data \cite{Bolliet:2015raa} but this conclusion clearly relies on heavy assumptions that might be radically altered when considering trans-planckian effects \cite{Martineau:2017tdx} or other ways of setting initial conditions \cite{Mielczarek:2014kea,Bojowald:2015gra}.\\

\begin{figure}[!h]
\begin{center}
\includegraphics[scale=0.38]{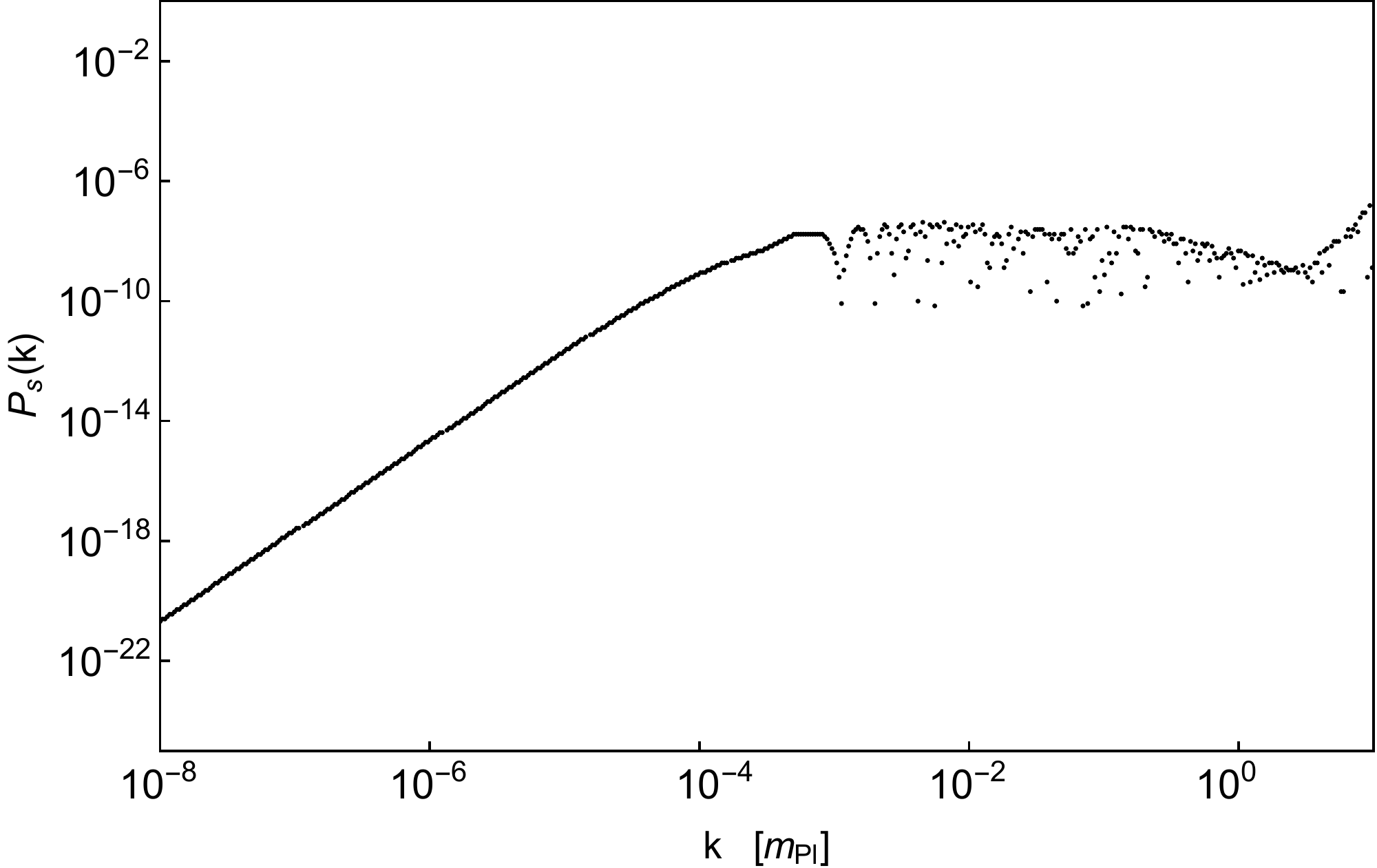}
\caption{Spectrum for $n = \frac{4}{3}$ within the deformed algebra approach and initial conditions set at $t_{i}=1.87\times10^{7}$.}
\label{da1}
\end{center}
\end{figure}

\begin{figure}[!h]
\begin{center}
\includegraphics[scale=0.38]{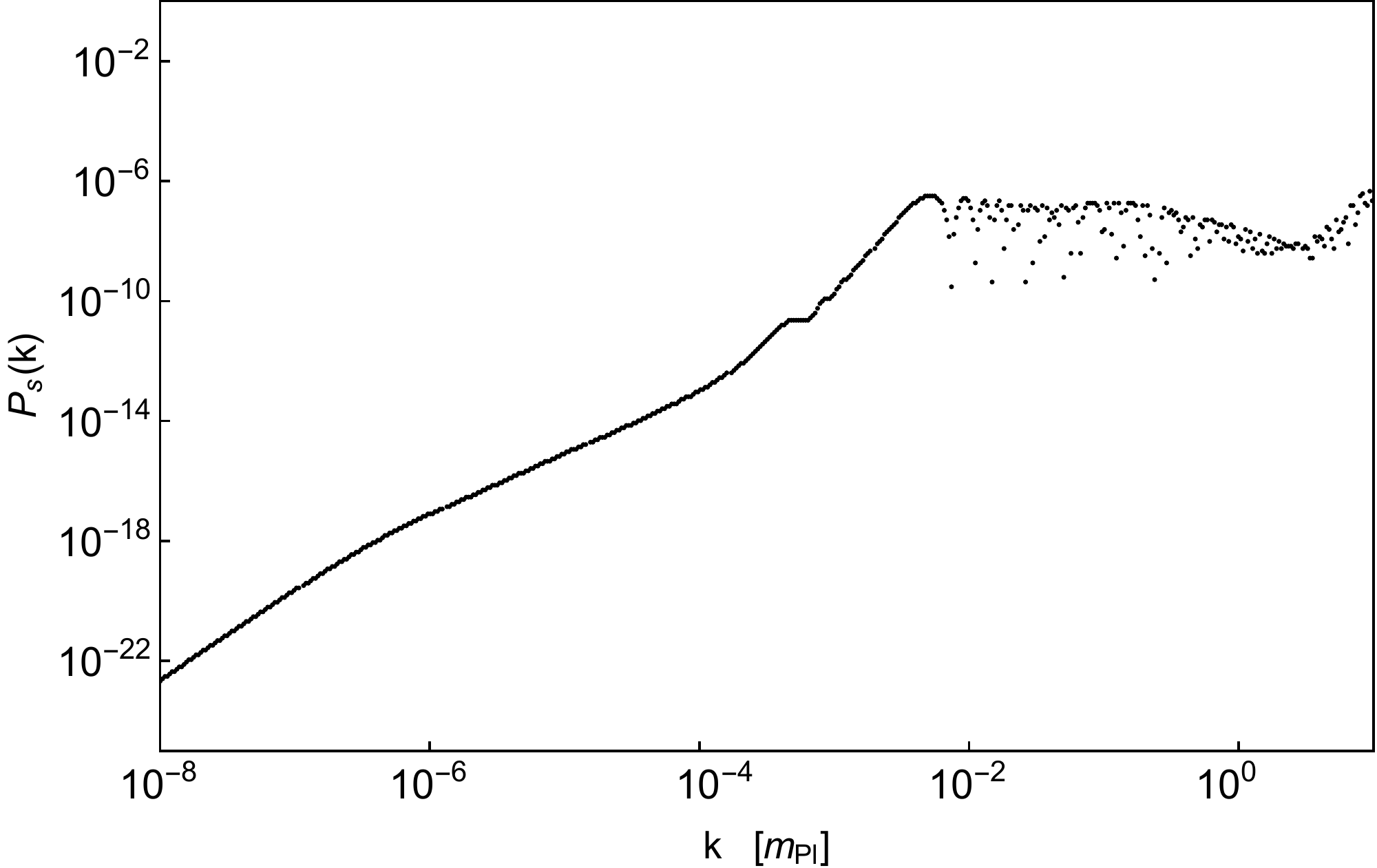}
\caption{Spectrum for $n = 3$ within the deformed algebra approach and initial conditions set at $t_{i}=1.17\times10^{7}$.}
\label{da2}
\end{center}
\end{figure}

\begin{figure}[!h]
\begin{center}
\includegraphics[scale=0.38]{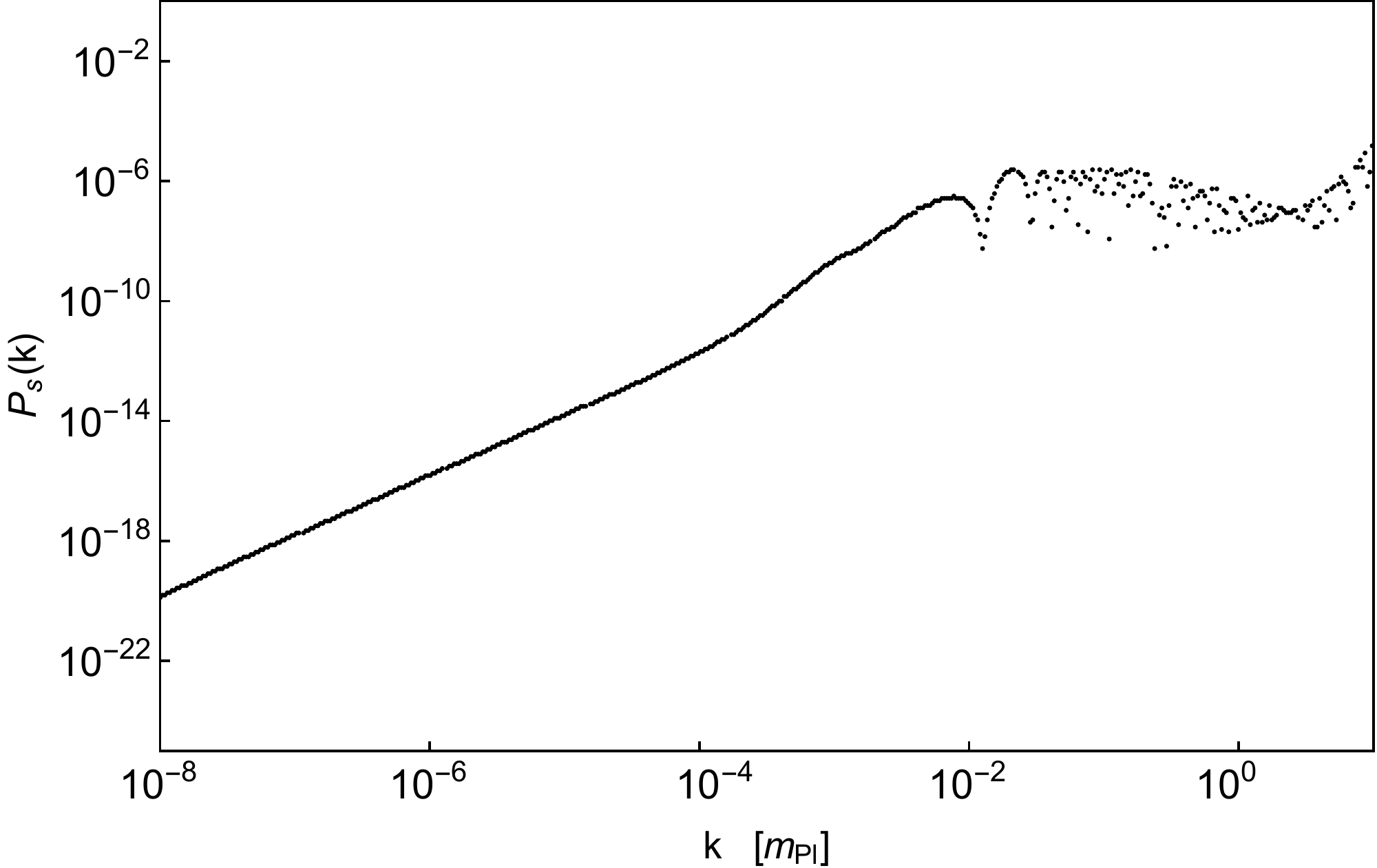}
\caption{Spectrum for $n = 4$ within the deformed algebra approach and initial conditions set at $t_{i}=1.19\times10^{7}$.}
\label{da3}
\end{center}
\end{figure}

We have readdressed the question of the propagation of scalar perturbations in the deformed algebra framework with new potentials. As it can be seen in Figs. \ref{da1},\ref{da2} and \ref{da3} the UV rise of the spectrum clearly remains present whatever the chosen potential. All the conclusions about the features of this model, therefore, remain valid beyond the massive scalar field approximation. The subtle modifications of the IR shape are actually due to the way the initial vacuum is chosen which is inevitably impacted by the choice of the potential.

\section{Conclusion}

In this article, we have addressed the question of the primordial power spectrum of scalar perturbations in a bouncing universe described by loop quantum cosmology by studying the gauge-invariant Mukhanov-Sasaki equation with the appropriate effective potential associated with different inflation potentials. A full numerical simulation was developed. The conclusions are the following:
\begin{itemize}
\item the temporal behavior of the effective $z''/z$ potential is, in general, highly complicated with a pseudo-periodic structure which depends on the details of the inflaton potential $V(\varphi )$.
\item the ultraviolet part of the power spectrum, which is the most relevant one from the observational perspective, is mostly independent of the way initial conditions are set and of the choice of the potential. This makes the main LQC predictions robust.
\item the intermediate and infrared parts of the spectrum do depend on the initial conditions and on the inflaton potential. The IR slope varies between $k^2$ and $k^3$ depending on the type of vacuum chosen and the amplitude of the oscillations can vary substantially.
\end{itemize}
This study shows that the main conclusions regarding the compatibility of the spectrum with CMB observations (for most of the parameter space span by initial conditions for the background) in LQC are reliable. However, if the initial values for the inflaton field and its momentum are fine-tuned so that the number of e-folds of inflation is small, the observational window might fall on the intermediate or IR part of the spectrum. In that case, LQC predictions do depend on the way initial conditions (for perturbations) are set and on the choice of the inflaton potential. This should be taken into account in future studies. 

\section*{Acknowledgments}

K.M is supported by a grant from the CFM foundation.

\bibliography{refs}

\begin{thebibliography}{48}
\expandafter\ifx\csname natexlab\endcsname\relax\def\natexlab#1{#1}\fi
\expandafter\ifx\csname bibnamefont\endcsname\relax
  \def\bibnamefont#1{#1}\fi
\expandafter\ifx\csname bibfnamefont\endcsname\relax
  \def\bibfnamefont#1{#1}\fi
\expandafter\ifx\csname citenamefont\endcsname\relax
  \def\citenamefont#1{#1}\fi
\expandafter\ifx\csname url\endcsname\relax
  \def\url#1{\texttt{#1}}\fi
\expandafter\ifx\csname urlprefix\endcsname\relax\def\urlprefix{URL }\fi
\providecommand{\bibinfo}[2]{#2}
\providecommand{\eprint}[2][]{\url{#2}}

\bibitem[{\citenamefont{Barrau}(2017)}]{Barrau:2017tcd}
\bibinfo{author}{\bibfnamefont{A.}~\bibnamefont{Barrau}},
  \bibinfo{journal}{Comptes Rendus Physique} \textbf{\bibinfo{volume}{18}},
  \bibinfo{pages}{189} (\bibinfo{year}{2017}), \eprint{1705.01597}.

\bibitem[{\citenamefont{Rovelli and Vidotto}(2014)}]{Rovelli:2014ssa}
\bibinfo{author}{\bibfnamefont{C.}~\bibnamefont{Rovelli}} \bibnamefont{and}
  \bibinfo{author}{\bibfnamefont{F.}~\bibnamefont{Vidotto}},
  \emph{\bibinfo{title}{{Covariant Loop Quantum Gravity}}}, Cambridge
  Monographs on Mathematical Physics (\bibinfo{publisher}{Cambridge University
  Press}, \bibinfo{year}{2014}), ISBN \bibinfo{isbn}{1107069629, 9781107069626,
  9781316147290}.

\bibitem[{\citenamefont{Ashtekar and Singh}(2011)}]{lqc9}
\bibinfo{author}{\bibfnamefont{A.}~\bibnamefont{Ashtekar}} \bibnamefont{and}
  \bibinfo{author}{\bibfnamefont{P.}~\bibnamefont{Singh}},
  \bibinfo{journal}{Class. Quant. Grav.} \textbf{\bibinfo{volume}{28}},
  \bibinfo{pages}{213001} (\bibinfo{year}{2011}), \eprint{1108.0893}.

\bibitem[{\citenamefont{Ashtekar and Barrau}(2015)}]{Ashtekar:2015dja}
\bibinfo{author}{\bibfnamefont{A.}~\bibnamefont{Ashtekar}} \bibnamefont{and}
  \bibinfo{author}{\bibfnamefont{A.}~\bibnamefont{Barrau}},
  \bibinfo{journal}{Class. Quant. Grav.} \textbf{\bibinfo{volume}{32}},
  \bibinfo{pages}{234001} (\bibinfo{year}{2015}), \eprint{1504.07559}.

\bibitem[{\citenamefont{Gerhardt et~al.}(2018)\citenamefont{Gerhardt, Oriti,
  and Wilson-Ewing}}]{Gerhardt:2018byq}
\bibinfo{author}{\bibfnamefont{F.}~\bibnamefont{Gerhardt}},
  \bibinfo{author}{\bibfnamefont{D.}~\bibnamefont{Oriti}}, \bibnamefont{and}
  \bibinfo{author}{\bibfnamefont{E.}~\bibnamefont{Wilson-Ewing}}
  (\bibinfo{year}{2018}), \eprint{1805.03099}.

\bibitem[{\citenamefont{Oriti}(2017)}]{Oriti:2016acw}
\bibinfo{author}{\bibfnamefont{D.}~\bibnamefont{Oriti}},
  \bibinfo{journal}{Comptes Rendus Physique} \textbf{\bibinfo{volume}{18}},
  \bibinfo{pages}{235} (\bibinfo{year}{2017}), \eprint{1612.09521}.

\bibitem[{\citenamefont{Oriti et~al.}(2017)\citenamefont{Oriti, Sindoni, and
  Wilson-Ewing}}]{Oriti:2016ueo}
\bibinfo{author}{\bibfnamefont{D.}~\bibnamefont{Oriti}},
  \bibinfo{author}{\bibfnamefont{L.}~\bibnamefont{Sindoni}}, \bibnamefont{and}
  \bibinfo{author}{\bibfnamefont{E.}~\bibnamefont{Wilson-Ewing}},
  \bibinfo{journal}{Class. Quant. Grav.} \textbf{\bibinfo{volume}{34}},
  \bibinfo{pages}{04LT01} (\bibinfo{year}{2017}), \eprint{1602.08271}.

\bibitem[{\citenamefont{Alesci et~al.}(2017)\citenamefont{Alesci, Botta,
  Cianfrani, and Liberati}}]{Alesci:2016xqa}
\bibinfo{author}{\bibfnamefont{E.}~\bibnamefont{Alesci}},
  \bibinfo{author}{\bibfnamefont{G.}~\bibnamefont{Botta}},
  \bibinfo{author}{\bibfnamefont{F.}~\bibnamefont{Cianfrani}},
  \bibnamefont{and} \bibinfo{author}{\bibfnamefont{S.}~\bibnamefont{Liberati}},
  \bibinfo{journal}{Phys. Rev.} \textbf{\bibinfo{volume}{D96}},
  \bibinfo{pages}{046008} (\bibinfo{year}{2017}), \eprint{1612.07116}.

\bibitem[{\citenamefont{Alesci and Cianfrani}(2016)}]{Alesci:2016gub}
\bibinfo{author}{\bibfnamefont{E.}~\bibnamefont{Alesci}} \bibnamefont{and}
  \bibinfo{author}{\bibfnamefont{F.}~\bibnamefont{Cianfrani}},
  \bibinfo{journal}{Int. J. Mod. Phys.} \textbf{\bibinfo{volume}{D25}},
  \bibinfo{pages}{1642005} (\bibinfo{year}{2016}), \eprint{1602.05475}.

\bibitem[{\citenamefont{Alesci and Cianfrani}(2015)}]{Alesci:2015nja}
\bibinfo{author}{\bibfnamefont{E.}~\bibnamefont{Alesci}} \bibnamefont{and}
  \bibinfo{author}{\bibfnamefont{F.}~\bibnamefont{Cianfrani}},
  \bibinfo{journal}{Phys. Rev.} \textbf{\bibinfo{volume}{D92}},
  \bibinfo{pages}{084065} (\bibinfo{year}{2015}), \eprint{1506.07835}.

\bibitem[{\citenamefont{Alesci and Cianfrani}(2013)}]{Alesci:2013xd}
\bibinfo{author}{\bibfnamefont{E.}~\bibnamefont{Alesci}} \bibnamefont{and}
  \bibinfo{author}{\bibfnamefont{F.}~\bibnamefont{Cianfrani}},
  \bibinfo{journal}{Phys. Rev.} \textbf{\bibinfo{volume}{D87}},
  \bibinfo{pages}{083521} (\bibinfo{year}{2013}), \eprint{1301.2245}.

\bibitem[{\citenamefont{Dapor and Liegener}(2018)}]{Dapor:2018sgb}
\bibinfo{author}{\bibfnamefont{A.}~\bibnamefont{Dapor}} \bibnamefont{and}
  \bibinfo{author}{\bibfnamefont{K.}~\bibnamefont{Liegener}},
  \bibinfo{journal}{Class. Quant. Grav.} \textbf{\bibinfo{volume}{35}},
  \bibinfo{pages}{135011} (\bibinfo{year}{2018}).

\bibitem[{\citenamefont{Engle and Vilensky}(2018)}]{Engle:2018zbe}
\bibinfo{author}{\bibfnamefont{J.}~\bibnamefont{Engle}} \bibnamefont{and}
  \bibinfo{author}{\bibfnamefont{I.}~\bibnamefont{Vilensky}}
  (\bibinfo{year}{2018}), \eprint{1802.01543}.

\bibitem[{\citenamefont{Wilson-Ewing}(2018)}]{Wilson-Ewing:2017vju}
\bibinfo{author}{\bibfnamefont{E.}~\bibnamefont{Wilson-Ewing}},
  \bibinfo{journal}{Class. Quant. Grav.} \textbf{\bibinfo{volume}{35}},
  \bibinfo{pages}{065005} (\bibinfo{year}{2018}), \eprint{1711.10943}.

\bibitem[{\citenamefont{Fernandez-Mendez
  et~al.}(2013)\citenamefont{Fernandez-Mendez, Mena~Marugan, and
  Olmedo}}]{Fernandez-Mendez:2013jqa}
\bibinfo{author}{\bibfnamefont{M.}~\bibnamefont{Fernandez-Mendez}},
  \bibinfo{author}{\bibfnamefont{G.~A.} \bibnamefont{Mena~Marugan}},
  \bibnamefont{and} \bibinfo{author}{\bibfnamefont{J.}~\bibnamefont{Olmedo}},
  \bibinfo{journal}{Phys. Rev.} \textbf{\bibinfo{volume}{D88}},
  \bibinfo{pages}{044013} (\bibinfo{year}{2013}), \eprint{1307.5222}.

\bibitem[{\citenamefont{Martin-de Blas et~al.}(2014)\citenamefont{Martin-de
  Blas, Martin-Benito, and Mena~Marugan}}]{Blas:2014naa}
\bibinfo{author}{\bibfnamefont{D.}~\bibnamefont{Martin-de Blas}},
  \bibinfo{author}{\bibfnamefont{M.}~\bibnamefont{Martin-Benito}},
  \bibnamefont{and} \bibinfo{author}{\bibfnamefont{G.~A.}
  \bibnamefont{Mena~Marugan}}, \bibinfo{journal}{Springer Proc. Math. Stat.}
  \textbf{\bibinfo{volume}{60}}, \bibinfo{pages}{327} (\bibinfo{year}{2014}).

\bibitem[{\citenamefont{Agullo et~al.}(2013{\natexlab{a}})\citenamefont{Agullo,
  Ashtekar, and Nelson}}]{Agullo1}
\bibinfo{author}{\bibfnamefont{I.}~\bibnamefont{Agullo}},
  \bibinfo{author}{\bibfnamefont{A.}~\bibnamefont{Ashtekar}}, \bibnamefont{and}
  \bibinfo{author}{\bibfnamefont{W.}~\bibnamefont{Nelson}},
  \bibinfo{journal}{Class.Quant.Grav.} \textbf{\bibinfo{volume}{30}},
  \bibinfo{pages}{085014} (\bibinfo{year}{2013}{\natexlab{a}}),
  \eprint{1302.0254}.

\bibitem[{\citenamefont{Agullo et~al.}(2012)\citenamefont{Agullo, Ashtekar, and
  Nelson}}]{Agullo2}
\bibinfo{author}{\bibfnamefont{I.}~\bibnamefont{Agullo}},
  \bibinfo{author}{\bibfnamefont{A.}~\bibnamefont{Ashtekar}}, \bibnamefont{and}
  \bibinfo{author}{\bibfnamefont{W.}~\bibnamefont{Nelson}},
  \bibinfo{journal}{Phys. Rev. Lett.} \textbf{\bibinfo{volume}{109}},
  \bibinfo{pages}{251301} (\bibinfo{year}{2012}), \eprint{1209.1609}.

\bibitem[{\citenamefont{Agullo et~al.}(2013{\natexlab{b}})\citenamefont{Agullo,
  Ashtekar, and Nelson}}]{Agullo3}
\bibinfo{author}{\bibfnamefont{I.}~\bibnamefont{Agullo}},
  \bibinfo{author}{\bibfnamefont{A.}~\bibnamefont{Ashtekar}}, \bibnamefont{and}
  \bibinfo{author}{\bibfnamefont{W.}~\bibnamefont{Nelson}},
  \bibinfo{journal}{Phys. Rev.} \textbf{\bibinfo{volume}{D87}},
  \bibinfo{pages}{043507} (\bibinfo{year}{2013}{\natexlab{b}}),
  \eprint{1211.1354}.

\bibitem[{\citenamefont{Mielczarek et~al.}(2012)\citenamefont{Mielczarek,
  Cailleteau, Barrau, and Grain}}]{tom1}
\bibinfo{author}{\bibfnamefont{J.}~\bibnamefont{Mielczarek}},
  \bibinfo{author}{\bibfnamefont{T.}~\bibnamefont{Cailleteau}},
  \bibinfo{author}{\bibfnamefont{A.}~\bibnamefont{Barrau}}, \bibnamefont{and}
  \bibinfo{author}{\bibfnamefont{J.}~\bibnamefont{Grain}},
  \bibinfo{journal}{Class. Quant. Grav.} \textbf{\bibinfo{volume}{29}},
  \bibinfo{pages}{085009} (\bibinfo{year}{2012}), \eprint{1106.3744}.

\bibitem[{\citenamefont{Cailleteau
  et~al.}(2012{\natexlab{a}})\citenamefont{Cailleteau, Mielczarek, Barrau, and
  Grain}}]{tom2}
\bibinfo{author}{\bibfnamefont{T.}~\bibnamefont{Cailleteau}},
  \bibinfo{author}{\bibfnamefont{J.}~\bibnamefont{Mielczarek}},
  \bibinfo{author}{\bibfnamefont{A.}~\bibnamefont{Barrau}}, \bibnamefont{and}
  \bibinfo{author}{\bibfnamefont{J.}~\bibnamefont{Grain}},
  \bibinfo{journal}{Class. Quant. Grav.} \textbf{\bibinfo{volume}{29}},
  \bibinfo{pages}{095010} (\bibinfo{year}{2012}{\natexlab{a}}),
  \eprint{1111.3535}.

\bibitem[{\citenamefont{Cailleteau
  et~al.}(2012{\natexlab{b}})\citenamefont{Cailleteau, Barrau, Grain, and
  Vidotto}}]{eucl2}
\bibinfo{author}{\bibfnamefont{T.}~\bibnamefont{Cailleteau}},
  \bibinfo{author}{\bibfnamefont{A.}~\bibnamefont{Barrau}},
  \bibinfo{author}{\bibfnamefont{J.}~\bibnamefont{Grain}}, \bibnamefont{and}
  \bibinfo{author}{\bibfnamefont{F.}~\bibnamefont{Vidotto}},
  \bibinfo{journal}{Phys. Rev.} \textbf{\bibinfo{volume}{D86}},
  \bibinfo{pages}{087301} (\bibinfo{year}{2012}{\natexlab{b}}),
  \eprint{1206.6736}.

\bibitem[{\citenamefont{Barrau et~al.}(2015)\citenamefont{Barrau, Bojowald,
  Calcagni, Grain, and Kagan}}]{eucl3}
\bibinfo{author}{\bibfnamefont{A.}~\bibnamefont{Barrau}},
  \bibinfo{author}{\bibfnamefont{M.}~\bibnamefont{Bojowald}},
  \bibinfo{author}{\bibfnamefont{G.}~\bibnamefont{Calcagni}},
  \bibinfo{author}{\bibfnamefont{J.}~\bibnamefont{Grain}}, \bibnamefont{and}
  \bibinfo{author}{\bibfnamefont{M.}~\bibnamefont{Kagan}},
  \bibinfo{journal}{JCAP} \textbf{\bibinfo{volume}{1505}}, \bibinfo{pages}{051}
  (\bibinfo{year}{2015}), \eprint{1404.1018}.

\bibitem[{\citenamefont{Agullo and Morris}(2015)}]{Agullo:2015tca}
\bibinfo{author}{\bibfnamefont{I.}~\bibnamefont{Agullo}} \bibnamefont{and}
  \bibinfo{author}{\bibfnamefont{N.~A.} \bibnamefont{Morris}},
  \bibinfo{journal}{Phys. Rev.} \textbf{\bibinfo{volume}{D92}},
  \bibinfo{pages}{124040} (\bibinfo{year}{2015}), \eprint{1509.05693}.

\bibitem[{\citenamefont{Bolliet et~al.}(2015)\citenamefont{Bolliet, Grain,
  Stahl, Linsefors, and Barrau}}]{Bolliet:2015bka}
\bibinfo{author}{\bibfnamefont{B.}~\bibnamefont{Bolliet}},
  \bibinfo{author}{\bibfnamefont{J.}~\bibnamefont{Grain}},
  \bibinfo{author}{\bibfnamefont{C.}~\bibnamefont{Stahl}},
  \bibinfo{author}{\bibfnamefont{L.}~\bibnamefont{Linsefors}},
  \bibnamefont{and} \bibinfo{author}{\bibfnamefont{A.}~\bibnamefont{Barrau}},
  \bibinfo{journal}{Phys.Rev.} \textbf{\bibinfo{volume}{D91}},
  \bibinfo{pages}{084035} (\bibinfo{year}{2015}), \eprint{1502.02431}.

\bibitem[{\citenamefont{Linsefors et~al.}(2013)\citenamefont{Linsefors,
  Cailleteau, Barrau, and Grain}}]{lcbg}
\bibinfo{author}{\bibfnamefont{L.}~\bibnamefont{Linsefors}},
  \bibinfo{author}{\bibfnamefont{T.}~\bibnamefont{Cailleteau}},
  \bibinfo{author}{\bibfnamefont{A.}~\bibnamefont{Barrau}}, \bibnamefont{and}
  \bibinfo{author}{\bibfnamefont{J.}~\bibnamefont{Grain}},
  \bibinfo{journal}{Phys. Rev.} \textbf{\bibinfo{volume}{D87}},
  \bibinfo{pages}{107503} (\bibinfo{year}{2013}), \eprint{1212.2852}.

\bibitem[{\citenamefont{Schander et~al.}(2016)\citenamefont{Schander, Barrau,
  Bolliet, Linsefors, Mielczarek, and Grain}}]{Schander:2015eja}
\bibinfo{author}{\bibfnamefont{S.}~\bibnamefont{Schander}},
  \bibinfo{author}{\bibfnamefont{A.}~\bibnamefont{Barrau}},
  \bibinfo{author}{\bibfnamefont{B.}~\bibnamefont{Bolliet}},
  \bibinfo{author}{\bibfnamefont{L.}~\bibnamefont{Linsefors}},
  \bibinfo{author}{\bibfnamefont{J.}~\bibnamefont{Mielczarek}},
  \bibnamefont{and} \bibinfo{author}{\bibfnamefont{J.}~\bibnamefont{Grain}},
  \bibinfo{journal}{Phys. Rev.} \textbf{\bibinfo{volume}{D93}},
  \bibinfo{pages}{023531} (\bibinfo{year}{2016}), \eprint{1508.06786}.

\bibitem[{\citenamefont{Martineau et~al.}(2018)\citenamefont{Martineau, Barrau,
  and Grain}}]{Martineau:2017tdx}
\bibinfo{author}{\bibfnamefont{K.}~\bibnamefont{Martineau}},
  \bibinfo{author}{\bibfnamefont{A.}~\bibnamefont{Barrau}}, \bibnamefont{and}
  \bibinfo{author}{\bibfnamefont{J.}~\bibnamefont{Grain}},
  \bibinfo{journal}{Int. J. Mod. Phys.} \textbf{\bibinfo{volume}{D27}},
  \bibinfo{pages}{1850067} (\bibinfo{year}{2018}), \eprint{1709.03301}.

\bibitem[{\citenamefont{Bolliet et~al.}(2016)\citenamefont{Bolliet, Barrau,
  Grain, and Schander}}]{Bolliet:2015raa}
\bibinfo{author}{\bibfnamefont{B.}~\bibnamefont{Bolliet}},
  \bibinfo{author}{\bibfnamefont{A.}~\bibnamefont{Barrau}},
  \bibinfo{author}{\bibfnamefont{J.}~\bibnamefont{Grain}}, \bibnamefont{and}
  \bibinfo{author}{\bibfnamefont{S.}~\bibnamefont{Schander}},
  \bibinfo{journal}{Phys. Rev.} \textbf{\bibinfo{volume}{D93}},
  \bibinfo{pages}{124011} (\bibinfo{year}{2016}), \eprint{1510.08766}.

\bibitem[{\citenamefont{Wilson-Ewing}(2012)}]{ed}
\bibinfo{author}{\bibfnamefont{E.}~\bibnamefont{Wilson-Ewing}},
  \bibinfo{journal}{Class. Quant. Grav.} \textbf{\bibinfo{volume}{29}},
  \bibinfo{pages}{215013} (\bibinfo{year}{2012}), \eprint{1205.3370}.

\bibitem[{\citenamefont{Gielen and Oriti}(2017)}]{Gielen:2017eco}
\bibinfo{author}{\bibfnamefont{S.}~\bibnamefont{Gielen}} \bibnamefont{and}
  \bibinfo{author}{\bibfnamefont{D.}~\bibnamefont{Oriti}}
  (\bibinfo{year}{2017}), \eprint{1709.01095}.

\bibitem[{\citenamefont{Agullo}(2018)}]{Agullo:2018wbf}
\bibinfo{author}{\bibfnamefont{I.}~\bibnamefont{Agullo}}
  (\bibinfo{year}{2018}), \eprint{1805.11356}.

\bibitem[{\citenamefont{Wilson-Ewing}(2017)}]{Wilson-Ewing:2016yan}
\bibinfo{author}{\bibfnamefont{E.}~\bibnamefont{Wilson-Ewing}},
  \bibinfo{journal}{Comptes Rendus Physique} \textbf{\bibinfo{volume}{18}},
  \bibinfo{pages}{207} (\bibinfo{year}{2017}), \eprint{1612.04551}.

\bibitem[{\citenamefont{Elizaga~Navascus
  et~al.}(2018)\citenamefont{Elizaga~Navascus, Martin~de Blas, and
  Mena~Marugn}}]{ElizagaNavascues:2017avq}
\bibinfo{author}{\bibfnamefont{B.}~\bibnamefont{Elizaga~Navascus}},
  \bibinfo{author}{\bibfnamefont{D.}~\bibnamefont{Martin~de Blas}},
  \bibnamefont{and} \bibinfo{author}{\bibfnamefont{G.~A.}
  \bibnamefont{Mena~Marugn}}, \bibinfo{journal}{Phys. Rev.}
  \textbf{\bibinfo{volume}{D97}}, \bibinfo{pages}{043523}
  (\bibinfo{year}{2018}), \eprint{1711.10861}.

\bibitem[{\citenamefont{Bonga and Gupt}(2016{\natexlab{a}})}]{Bonga:2015kaa}
\bibinfo{author}{\bibfnamefont{B.}~\bibnamefont{Bonga}} \bibnamefont{and}
  \bibinfo{author}{\bibfnamefont{B.}~\bibnamefont{Gupt}},
  \bibinfo{journal}{Gen. Rel. Grav.} \textbf{\bibinfo{volume}{48}},
  \bibinfo{pages}{71} (\bibinfo{year}{2016}{\natexlab{a}}),
  \eprint{1510.00680}.

\bibitem[{\citenamefont{Bonga and Gupt}(2016{\natexlab{b}})}]{Bonga:2015xna}
\bibinfo{author}{\bibfnamefont{B.}~\bibnamefont{Bonga}} \bibnamefont{and}
  \bibinfo{author}{\bibfnamefont{B.}~\bibnamefont{Gupt}},
  \bibinfo{journal}{Phys. Rev.} \textbf{\bibinfo{volume}{D93}},
  \bibinfo{pages}{063513} (\bibinfo{year}{2016}{\natexlab{b}}),
  \eprint{1510.04896}.

\bibitem[{\citenamefont{Agullo et~al.}(2017)\citenamefont{Agullo, Ashtekar, and
  Gupt}}]{Agullo:2016hap}
\bibinfo{author}{\bibfnamefont{I.}~\bibnamefont{Agullo}},
  \bibinfo{author}{\bibfnamefont{A.}~\bibnamefont{Ashtekar}}, \bibnamefont{and}
  \bibinfo{author}{\bibfnamefont{B.}~\bibnamefont{Gupt}},
  \bibinfo{journal}{Class. Quant. Grav.} \textbf{\bibinfo{volume}{34}},
  \bibinfo{pages}{074003} (\bibinfo{year}{2017}), \eprint{1611.09810}.

\bibitem[{\citenamefont{Ashtekar and Gupt}(2017)}]{Ashtekar:2016pqn}
\bibinfo{author}{\bibfnamefont{A.}~\bibnamefont{Ashtekar}} \bibnamefont{and}
  \bibinfo{author}{\bibfnamefont{B.}~\bibnamefont{Gupt}},
  \bibinfo{journal}{Class. Quant. Grav.} \textbf{\bibinfo{volume}{34}},
  \bibinfo{pages}{035004} (\bibinfo{year}{2017}), \eprint{1610.09424}.

\bibitem[{\citenamefont{Linsefors and Barrau}(2013)}]{bl}
\bibinfo{author}{\bibfnamefont{L.}~\bibnamefont{Linsefors}} \bibnamefont{and}
  \bibinfo{author}{\bibfnamefont{A.}~\bibnamefont{Barrau}},
  \bibinfo{journal}{Phys. Rev.} \textbf{\bibinfo{volume}{D87}},
  \bibinfo{pages}{123509} (\bibinfo{year}{2013}), \eprint{1301.1264}.

\bibitem[{\citenamefont{Linsefors and Barrau}(2015)}]{Linsefors:2014tna}
\bibinfo{author}{\bibfnamefont{L.}~\bibnamefont{Linsefors}} \bibnamefont{and}
  \bibinfo{author}{\bibfnamefont{A.}~\bibnamefont{Barrau}},
  \bibinfo{journal}{Class. Quant. Grav.} \textbf{\bibinfo{volume}{32}},
  \bibinfo{pages}{035010} (\bibinfo{year}{2015}), \eprint{1405.1753}.

\bibitem[{\citenamefont{Martineau et~al.}(2017)\citenamefont{Martineau, Barrau,
  and Schander}}]{Martineau:2017sti}
\bibinfo{author}{\bibfnamefont{K.}~\bibnamefont{Martineau}},
  \bibinfo{author}{\bibfnamefont{A.}~\bibnamefont{Barrau}}, \bibnamefont{and}
  \bibinfo{author}{\bibfnamefont{S.}~\bibnamefont{Schander}},
  \bibinfo{journal}{Phys. Rev.} \textbf{\bibinfo{volume}{D95}},
  \bibinfo{pages}{083507} (\bibinfo{year}{2017}), \eprint{1701.02703}.

\bibitem[{\citenamefont{Barrau and Bolliet}(2016)}]{Barrau:2016nwy}
\bibinfo{author}{\bibfnamefont{A.}~\bibnamefont{Barrau}} \bibnamefont{and}
  \bibinfo{author}{\bibfnamefont{B.}~\bibnamefont{Bolliet}},
  \bibinfo{journal}{Int. J. Mod. Phys.} \textbf{\bibinfo{volume}{D25}},
  \bibinfo{pages}{1642008} (\bibinfo{year}{2016}), \eprint{1602.04452}.

\bibitem[{\citenamefont{Espinoza-Garca and
  Torres-Lomas}(2017)}]{Espinoza-Garcia:2017qjl}
\bibinfo{author}{\bibfnamefont{A.}~\bibnamefont{Espinoza-Garca}}
  \bibnamefont{and}
  \bibinfo{author}{\bibfnamefont{E.}~\bibnamefont{Torres-Lomas}}
  (\bibinfo{year}{2017}), \eprint{1709.03242}.

\bibitem[{\citenamefont{Ade et~al.}(2016)}]{Ade:2015lrj}
\bibinfo{author}{\bibfnamefont{P.~A.~R.} \bibnamefont{Ade}}
  \bibnamefont{et~al.} (\bibinfo{collaboration}{Planck}),
  \bibinfo{journal}{Astron. Astrophys.} \textbf{\bibinfo{volume}{594}},
  \bibinfo{pages}{A20} (\bibinfo{year}{2016}), \eprint{1502.02114}.

\bibitem[{\citenamefont{Bojowald and Paily}(2012)}]{Bojowald:2011aa}
\bibinfo{author}{\bibfnamefont{M.}~\bibnamefont{Bojowald}} \bibnamefont{and}
  \bibinfo{author}{\bibfnamefont{G.~M.} \bibnamefont{Paily}},
  \bibinfo{journal}{Phys. Rev.} \textbf{\bibinfo{volume}{D86}},
  \bibinfo{pages}{104018} (\bibinfo{year}{2012}), \eprint{1112.1899}.

\bibitem[{\citenamefont{Cailleteau et~al.}(2014)\citenamefont{Cailleteau,
  Linsefors, and Barrau}}]{Cailleteau:2013kqa}
\bibinfo{author}{\bibfnamefont{T.}~\bibnamefont{Cailleteau}},
  \bibinfo{author}{\bibfnamefont{L.}~\bibnamefont{Linsefors}},
  \bibnamefont{and} \bibinfo{author}{\bibfnamefont{A.}~\bibnamefont{Barrau}},
  \bibinfo{journal}{Class. Quant. Grav.} \textbf{\bibinfo{volume}{31}},
  \bibinfo{pages}{125011} (\bibinfo{year}{2014}), \eprint{1307.5238}.

\bibitem[{\citenamefont{Mielczarek et~al.}(2018)\citenamefont{Mielczarek,
  Linsefors, and Barrau}}]{Mielczarek:2014kea}
\bibinfo{author}{\bibfnamefont{J.}~\bibnamefont{Mielczarek}},
  \bibinfo{author}{\bibfnamefont{L.}~\bibnamefont{Linsefors}},
  \bibnamefont{and} \bibinfo{author}{\bibfnamefont{A.}~\bibnamefont{Barrau}},
  \bibinfo{journal}{Int. J. Mod. Phys.} \textbf{\bibinfo{volume}{D27}},
  \bibinfo{pages}{1850050} (\bibinfo{year}{2018}), \eprint{1411.0272}.

\bibitem[{\citenamefont{Bojowald and Mielczarek}(2015)}]{Bojowald:2015gra}
\bibinfo{author}{\bibfnamefont{M.}~\bibnamefont{Bojowald}} \bibnamefont{and}
  \bibinfo{author}{\bibfnamefont{J.}~\bibnamefont{Mielczarek}},
  \bibinfo{journal}{JCAP} \textbf{\bibinfo{volume}{1508}}, \bibinfo{pages}{052}
  (\bibinfo{year}{2015}), \eprint{1503.09154}.

\end{thebibliography}
\end{document}